\documentclass[aps,prc,superscriptaddress,showpacs,nofootinbib,twocolumn]{
revtex4-1}

\usepackage{graphicx,amssymb,amsmath,dcolumn,gensymb,hyperref}

\newcommand{\unit}[1]{\ensuremath{\, \mathrm{#1}}}
\newcommand{\etal}{\textit{et al.} }
\newcommand{\etaln}{\textit{et al.}}
\newcommand{\ie}{\textit{i.e.}, }
\newcommand{\boldq}{\ensuremath{\boldsymbol{q}}}

\newcommand{\boldn}{\ensuremath{\boldsymbol{n}}}
\newcolumntype{d}[1]{D{.}{.}{#1} }

\begin{document}

\title{Domains and defects in nuclear ``pasta''}
\author{A.~S. Schneider}\email{andschn@caltech.edu}
\affiliation{TAPIR, Walter Burke Institute for Theoretical Physics, MC 350-17,  
California Institute of Technology, Pasadena, California 91125, USA}
\author{M.~E. Caplan}\email{mecaplan@physics.mcgill.ca}
\affiliation{Department of Physics and McGill Space Institute, McGill 
University, 3600 rue University, Montreal QC, Canada H3A 2T8}
\author{D.~K. Berry}\email{dkberry@iu.edu}
\author{C.~J. Horowitz}\email{horowit@indiana.edu}
\affiliation{Center for Exploration of Energy and Matter and Department of  
Physics, Indiana University, Bloomington, IN 47405, USA}
\date{\today}
\begin{abstract}


Nuclear pasta topology is an essential ingredient to determine transport 
properties in the inner crust of neutron stars.
We perform semi-classical molecular dynamics simulations of nuclear pasta for 
proton fractions $Y_p=0.30$ and $Y_p=0.40$ near one third of nuclear 
saturation density, $n=0.05\unit{fm}^{-3}$, at a temperature $T=1.0\unit{MeV}$.
Our simulations are, to our knowledge, the largest nuclear pasta simulations to 
date and contain up to 3\,276\,800 nucleons in the  $Y_p=0.30$ and 819\,200 
nucleons in the $Y_p=0.40$ case. 
An algorithm to determine which nucleons are part of a given sub-domain in the 
system is presented. 
By comparing runs of different sizes we study finite size effects, 
equilibration time, the formation of multiple domains and defects in the pasta 
structures, as well as the structure factor dependence on simulation size. 
Although we find qualitative agreement between the topological 
structure and the structure factors of runs with 51\,200 nucleons and 
those with 819\,200 nucleons or more, we show that simulations with hundreds of 
thousands of nucleons may be necessary to accurately predict pasta 
transport properties.


\end{abstract}



\maketitle

\section{Introduction}\label{sec:Intro}

At the base of the crust of neutrons stars (NSs) there is a dense system of 
nucleons  immersed in a degenerate relativistic electron gas. 
Because of the high density, $\rho\sim10^{14}\unit{g/cm}^3$, and Pauli 
blocking the degenerate electrons have a relatively long mean free path. 
Thus, electron transport dominates the system's electrical conductivity, thermal 
conductivity, and shear viscosity \cite{nandkumar:84}, although neutrons may 
have a non-negligible contribution \cite{chugunov:05}. 
Electron transport properties depend mainly on how electrons interact with 
protons. 
At the high densities found in the crust of NSs protons and neutrons may
cluster into exotic non-spherical shapes known as nuclear pasta 
\cite{ravenhall:83, hashimoto:84, oyamatsu:84, williams:85}. 
Hence, nuclear pasta topology determines transport properties at the base of 
the inner crust of NSs \cite{horowitz:04a, horowitz:04b,  horowitz:05, 
horowitz:08, horowitz:15, schneider:14, schneider:16, nandi:18}. 
These exotic nuclear shapes also determine neutrino transport in non-trivial 
ways \cite{watanabe:01, sonoda:07, alloy:11, piekarewicz:12, alcain:14b, 
horowitz:15, schneider:16, furtado:16, alcain:17, nandi:18}, have an impact on 
core-collapse supernovae (CCSNe) \cite{ogasawara:82, ogasawara:85, newton:09, 
watanabe:09, horowitz:16, roggero:18}, influence the structure and evolution of 
NSs \cite{pethick:98, watanabe:00, magierski:02, gearheart:11, pons:13, 
pais:16a, pais:16b, passamonti:16} and their cooling curves \cite{newton:13, 
merritt:16, brown:17, cumming:17, deibel:17}, as well as affect the final 
state of matter ejected during binary neutron star mergers \cite{caplan:15, 
alcain:18}. 
Particularly, the presence of nuclear pasta may significantly alter the 
elastic properties of the inner crust of NSs. 
Thus, nuclear pasta may impact the lifetimes and size of mountains on NS 
crusts, which could produce continuous gravitational waves detectable by the 
Advanced LIGO and VIRGO detectors \cite{abbott:17}.
The elastic properties of nuclear pasta are the subject of a companion Letter 
\cite{caplan:18}.

Because nuclear pasta only exists under conditions achieved in the interior of 
NSs and during CCSNe, its existence can only be inferred through indirect means 
\cite{pons:13, horowitz:15, merritt:16, deibel:17} and its properties have to be 
studied via numerical simulations, such as molecular dynamics (MD). 
An overview of MD simualtions of nuclear pasta is presented by Caplan and 
Horowitz in Ref. \cite{caplan:17}. 
Nuclear pasta is sensitive to temperature, density and proton fraction of the  
system \cite{magierski:02, maruyama:98, maruyama:04, maruyama:05, 
dorso:12, schneider:13, dorso:18, alcain:14a, nandi:16, avancini:17, 
sagert:16} and to yet unconstrained properties of nuclear matter 
\cite{avancini:08, avancini:10, pais:10, pais:14, grill:12, bao:14, alam:17, 
fattoyev:17}. 
Despite plenty of investigations using several different approaches a 
phase-diagram of nuclear pasta \cite{watanabe:02, watanabe:03, sonoda:08, 
nandi:17, pais:12, pais:15, schuetrumpf:13, schuetrumpf:14} and all of its 
possible topologies \cite{watanabe:03, watanabe:04, okamoto:12, okamoto:13, 
nakazato:09, nakazato:11, schuetrumpf:15a, schuetrumpf:15b, kubis:17, kycia:17} 
is still elusive. 
Amongst the issue faced are that analytical computations are limited to few 
symmetries \cite{ravenhall:83, avancini:08, kycia:17, kubis:17}, numerical 
simulations that use simplified nucleon-nucleon potentials are constrained by 
finite size effects \cite{alcain:14a, schneider:14, schneider:16, molinelli:14, 
molinelli:15, gimenez:15, alcain:18}, while computational power is an 
impediment for detailed quantum approaches \cite{sagert:16, fattoyev:17}. 
Furthermore, strong magnetic fields such as the ones in NSs or produced 
during CCSNe may significantly alter the topology of the pasta 
\cite{kobyakov:17, delima:13, ofengeim:15, passamonti:16}, but are rarely 
taken into account.

Another interesting aspect of the pasta phases is that their topology has 
equivalents in Skyrmion systems \cite{milde:13, kawaguchi:18}, polymers 
\cite{seddon:95, marrink:04, horsch:05, fodor:17, lazutin:17, lopez:17}, as 
well as in biological systems \cite{terasaki:13, guven:14, berry:16}. 
Past work has used nuclear pasta simulations to make insights into the physics 
of systems at completely different scales, such as biophysical membranes in 
eukaryotic cells \cite{berry:16}. This suggests that the structures formed by 
these self assembling systems are not dependent on the exact details of the 
microscopic interactions; rather, it may be possible to explain these 
commonalities with some simple universal geometric arguments.

Numerical simulations of nuclear pasta that incorporate quantum mechanics are 
necessary to resolve detailed properties of the nucleons that make up the 
pasta. 
However, those calculations are computationally expensive and, to date, are  
limited to hundreds to a few thousand nucleons \cite{newton:09, 
pais:12, sagert:16, fattoyev:17}. 
Molecular dynamics (MD) simulations show that finite size effects and boundary  
conditions are important for such small simulations and influence the pasta 
shapes formed \cite{alcain:14a, gimenez:15, alcain:18}. 
Moreover, to determine transport properties of nuclear pasta simulations with 
hundreds of thousands of nucleons or even more may be necessary 
\cite{horowitz:04a, horowitz:05, schneider:14, nandi:18}.

In this manuscript we discuss, to our knowledge, the largest nuclear 
pasta simulations to date. 
Using the Indiana University Molecular Dynamics (IUMD) code \cite{berry:13, 
schneider:14, caplan:15}, we compare results for the topology and transport 
properties of nuclear pasta for simulations containing up to 3\,276\,800 
nucleons for proton fractions of $Y_p=0.30$ and $819\,200$ for $Y_p=0.40$. 
We also discuss a method to discriminate domains within the simulation volume 
and examine how these evolve over time. 
In Sec. \ref{sec:Formalism} we review our MD formalism, discuss code 
performance, as well as present our algorithms to compute structure factors 
from our MD simulations and to differentiate domains within the simulation 
volume. 
We present our results for $Y_p=0.30$ systems in Sec. \ref{sec:Results} and for 
$Y_p=0.40$ systems in Sec. \ref{ssec:yp40}. 
We conclude and discuss present challenges in Sec. \ref{sec:Conclusions}.

\section{Formalism}\label{sec:Formalism}

The formalism of our molecular dynamics (MD) study is the same of many previous 
works initiated by Horowitz \etal \cite{horowitz:04a}. 
For a review refer to Ref. \cite{caplan:17} and references therein. 
In our MD simulations we model nucleons as point-like particles immersed in 
a background electron gas. 
We consider $N$ nucleons, $N_p$ protons and $N_n$ neutrons such that 
$N=N_p+N_n$, inside cubic volumes of side $L$ with periodic boundary 
conditions. 
The number density of the system is $n=N/L^3$ while its proton fraction is 
$Y_p=N_p/N$. 
Nucleons interact via a two-body force limited to the the nearest periodic 
image of other nucleons. 
The interaction potential depends on nucleon isospins and their inter-particle 
distances $r$ and has the form
\begin{subequations}\label{eq:potential}
\begin{align}
 V_{np}(r)&=a e^{-r^2/\Lambda}+[b-c]e^{-r^2/2\Lambda},\\
 V_{nn}(r)&=a e^{-r^2/\Lambda}+[b+c]e^{-r^2/2\Lambda},\\
 V_{pp}(r)&=a e^{-r^2/\Lambda}+[b+c]e^{-r^2/2\Lambda} 
            + \frac{\alpha}{r}e^{-r/\lambda}.
\end{align}
\end{subequations}
The subscripts $n$ and $p$ denote, respectively, whether a nucleon is a neutron 
or a proton.
The parameters $a=110\unit{MeV}$, $b=-26\unit{MeV}$, $c=24\unit{MeV}$ and  
$\Lambda=1.25\unit{fm}^2$ were fit to reproduce some properties of finite 
nuclei, pure neutron matter and symmetric nuclear matter \cite{horowitz:04a}. 
The long-range Coulomb repulsion between protons is screened by the background  
electron gas which renders the system electrically neutral. 
The relativistic Thomas-Fermi screening length $\lambda$ is given by
\begin{equation}
\label{eq:lambda}
\lambda = \frac{\pi^{1/2}}{2\alpha^{1/2}} 
\left(k_F\sqrt{k_F^2+m_e^2}\right)^{-1/2}
\end{equation}
where $k_F=(3\pi^2n_e)^{1/3}$ is the Fermi momentum of electrons with density 
$n_e$ and mass $m_e$. 
For electrically neutral systems $n_e=Y_pn$. 
As in previous works we set the screening length $\lambda$ to $10\unit{fm}$. 
For the proton fractions considered in this work this value is somewhat 
shorter than the value obtained considering non-interacting relativistic 
electrons, Eq. \eqref{eq:lambda}. 
However, we do not expect this difference to significantly influence our 
results \cite{schneider:14,schneider:16}.

Using this formalism, we simulate twelve systems at a constant density $n = 
0.05\unit{fm}^{-3}$ and constant temperature $T=1\unit{MeV}$. 
Five of runs have proton fraction $Y_p=0.30$ and contain 51\,200, 409\,600, 
819\,200, 1\,638\,400, and 3\,276\,800 nucleons. 
These runs are discussed in Sec. \ref{ssec:yp30}. 
Seven runs have proton fraction $Y_p=0.40$, and contain 51\,200, 61\,440, 
76\,800, 102\,400, 204\,800, 409\,600, and 819\,200 nucleons.
These are discussed in Sec. \ref{ssec:yp40}.

\subsection{The IUMD code performance}

Our MD simulations were run on the Big Red 2 supercomputer at Indiana 
University and on the Titan supercomputer at Oak Ridge National Laboratory.
Runs were performed using the IUMD CPU/GPU hybrid code described in Refs. 
\cite{berry:13, schneider:14, caplan:15}. 
Appropriate understanding of code performance and its limitations play a large 
role in determining the feasibility of state-of-the-art runs. 
Tracking bottlenecks in the code and causes for performance variability and 
degradation over time is an important factor in optimizing the usage of 
resources. 
Therefore, here we review some code details and examine its performance for our 
simulations with proton fraction $Y_p=0.30$.

In the IUMD code short range nuclear forces are computed on CPUs using 
a neighbor list scheme where only nucleons within $11\unit{fm}$ of each 
other interact. 
Hence, computation of nuclear forces scales with $\mathcal{O}(N)$. 
Meanwhile, long range Coulomb interaction between protons is distributed 
across the GPUs and scales as $\mathcal{O}(Y_p^2N^2)$. 
For large systems, such as the ones considered in this work, the bulk of the 
computational time is spent computing the Coulomb force. 
Thus, a simulation time step should be, to a good approximation, proportional 
to $(Y_pN)^2/P$, where $Y_p$ is the proton fraction $N$ the number of nucleons 
and $P$ the number of CPU/GPU units used. 
For an early discussion on the code scalability using only Coulomb interactions 
we refer to Berry \etal \cite{berry:13}.

\begin{figure}[!htb]
\centering
\includegraphics[trim=30 10 0 0, clip, 
width=0.45\textwidth]{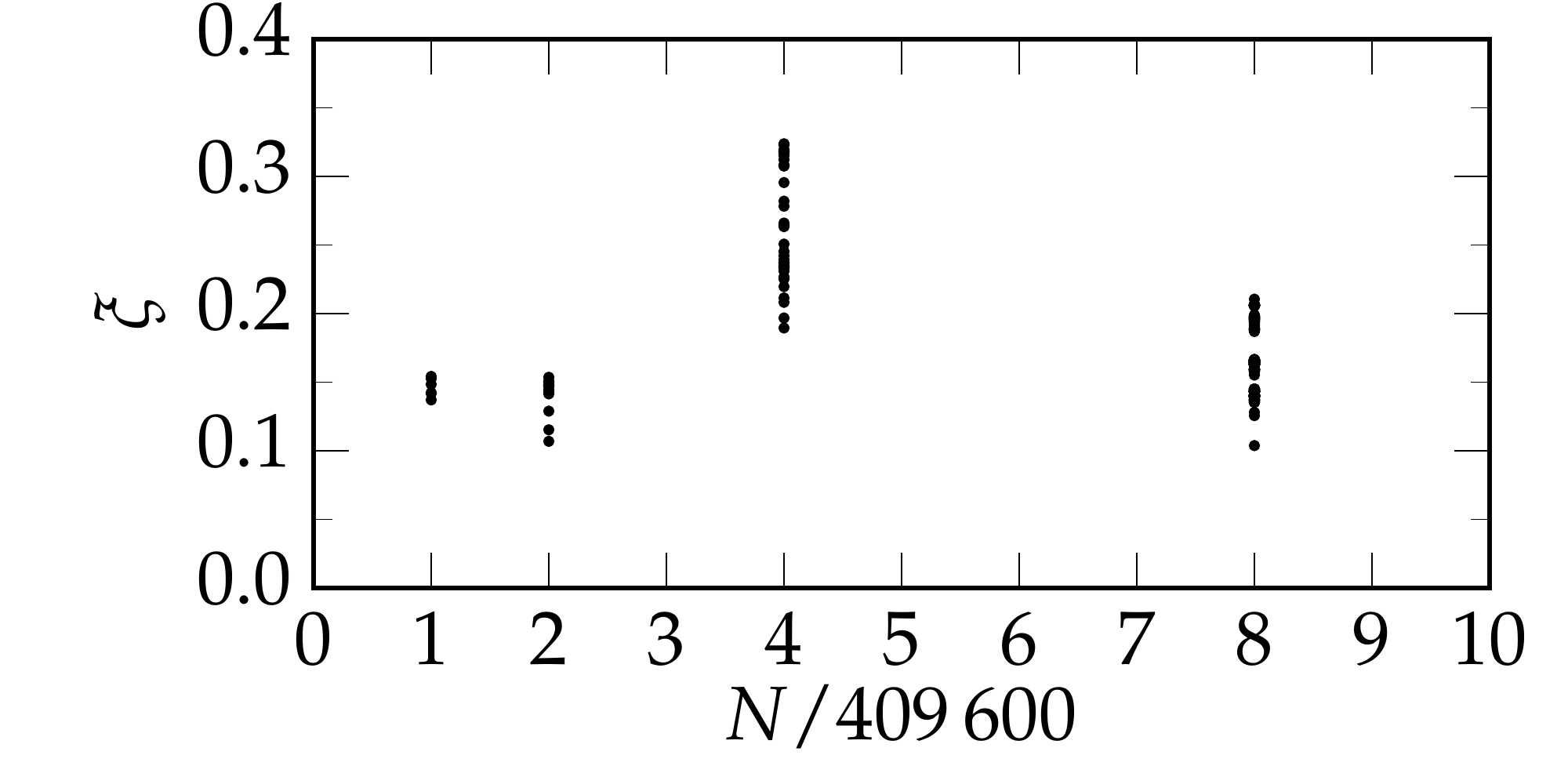}
\caption{\label{fig:perf} (Color online) Plot of the performance factor
$\xi = k_\xi(Y_pN)^2/(PT')$ for our runs with $N$ nucleons and 
proton fraction $Y_p=0.30$. 
The larger $\xi$ the better the performance. 
Note that $\xi$ is normalized to an arbitrary value as only ratios between 
different $\xi$s are meaningful.}
\end{figure}

In Fig. \ref{fig:perf} we plot the code performance $\xi=k_\xi(Y_pN)^2/(PT')$ 
for our $Y_p=0.30$ simulations with $N=409\,600$ nucleons as well as $2N$, 
$4N$, and $8N$ nucleons. 
Here, $k_\xi$ is a proportionality constant and $T'$ the real time necessary to 
compute a simulation time step. 
Ideally, the value of $\xi$ should remain constant across simulations of 
different sizes and throughout each run. 
However, we observe differences that depend on the simulation size, number 
of computing nodes used, and on the compiler version used to compile the IUMD 
code.

The 409\,800 nucleons run was performed on the hybrid CPU/GPU nodes 
of the Big Red 2 supercomputer. 
Within this run the number of CPU/GPU nodes was chosen to be either 64, 80, 
100, or 128. 
The entirety of this run was performed using the IUMD code compiled with PGI 
compiler version 14.1. 
We observed an approximate 5, 10, and 15\% decrease in code performance as the 
number of nodes requested were increased, respectively, from 64 to 80, 100, and 
128.

The runs with 819\,200 and 1\,638\,400 nucleons were performed exclusively 
using 1024 nodes on the hybrid CPU/GPU nodes of the Titan supercomputer. 
Although both Titan and Big Red 2 have NVIDIA Tesla GPU Accelerators, Titan is 
equipped with the K20X model while Big Red 2 is equipped with the K20 model. 
Thus, since Titan GPUs have a slightly larger number of CUDA cores (2688 to 
2496) and faster base clock (732 to 706 MHz) than Big Red 2 GPUs, this should 
translate into better performance. 
However, comparing the results of the 409\,800 and 819\,200 runs, we note 
that the performance of the smaller run is on average slightly better. 
This is likely due to the use of an excessive number of nodes in the 819\,200 
nucleon run, which degraded performance. 
This is clear from the fact that the 1\,638\,400 nucleon run performs 
significantly better than both the 409\,800 and the 819\,200 nucleon runs.

We noticed significant performance changes throughout the Titan runs, with 
$0.10\leq\xi\leq0.16$ for the 819\,200 nucleon run and $0.19\leq\xi\leq0.33$ for
the 1\,638\,400 nucleon run. 
In parts of the run the variability in performance could be tracked down 
to unusually slow MPI communication times between the requested nodes, while in 
other cases variability was due to small improvements in the IUMD code that 
were implemented during early stages of these two runs. 
However, the main aspect dictating the variability in code performance was the 
compiler version used. 
The IUMD code performs best when compiled with the PGI compiler version 14.1. 
This compiler version boosts the performance by at least 25\% when compared to 
any of the other eight PGI compiler versions tried, most of them more recent 
versions than 14.1.

Our largest run, with 3\,276\,800 nucleons, was performed exclusively on the 
Big Red 2 supercomputer. 
The number of nodes used was modified throughout the run and changed between 
64, 128, 256, and 512, depending on machine availability. 
Due to the large number of nucleons in this run, we did not observe changes in 
performance of more than 10\% by altering the number of nodes used from 64 to 
512, a factor of eight. 
Nevertheless, as was the case for the Titan runs, the most significant 
determinant in code performance was the PGI compiler version used. 
Yet again, the PGI CUDA Fortran compiler version 14.1 significantly outperformed 
any of the other versions tried.

\subsection{Structure factor}\label{ssec:sq}

Nuclear pasta shapes are relevant in astrophysical scenarios as they determine 
neutrino transport in core-collapse supernovae (CCSNe) and cooling time-scales 
of neutron stars (NSs) \cite{horowitz:04a, sonoda:08, newton:09, alloy:11, 
pais:12, horowitz:15, yakovlev:15, horowitz:16, nandi:18} as well as NS crust 
properties \cite{gearheart:11, pons:13, schneider:16}. 
Transport properties such as viscosity and electrical and heat conductivity are 
a function of the structure factor of the pasta shapes \cite{horowitz:04a, 
alcain:14b, schneider:16, nandi:18}. 
Thus, one of our interests is to compute if and how structure factors may be 
affected between simulations that differ by orders of magnitude in the number 
of nucleons.

At different points of our simulations we obtain a trajectory file of all 
nucleons by saving their positions every $200$ fm/$c$ for $10^6$ fm/$c$.
We use these trajectory files to determine nucleon structure factors  
following Refs. \cite{horowitz:08, schneider:13, schneider:14, schneider:16, 
nandi:18} and reviewed below.
The structure factor $S_i(\boldq)$ for a given transferred momentum 
$\boldq$ for a nucleon of type $i = n,\,p$ is given by the time average 
of the nucleon density in momentum space: 
\begin{equation}\label{eq:sq}
 S_i(\boldq) = 
 \langle\rho_i^\star(\boldq,t)\rho_i(\boldq,t)\rangle_t 
-\langle\rho_i^\star(\boldq,t)\rangle_t
 \langle\rho_i  (\boldq,t)\rangle_t.
\end{equation}
Above, $\rho_i(\boldq,t) = N_i^{-1/2} \sum_{j=1}^{N_i} 
e^{i\boldq \cdot \boldsymbol{r}_j(t)}$ is the nucleon density in 
momentum space, $\rho^\star(\boldq,t)$ its complex conjugate, with $N_i$ the 
number of nucleons of type $i$, $\boldsymbol{r}_j(t)$ the position of the $j$-th 
nucleon of type $i$ at time $t$, and the angled brackets $\langle{A}\rangle_a$ 
denote the average of quantity $A$ over a set of $a$. 
To avoid finite-size effects in the computations of $\rho_i(\boldq,t)$ 
due to the periodic boundary conditions imposed in the system we only take into 
account momenta $\boldq$ such that 
\begin{equation}\label{eq:q}
\boldq = 2\pi \left(\frac{n_x}{L_x}, \frac{n_y}{L_y}, 
\frac{n_z}{L_z}\right),
\end{equation}
where $n_i$ are integers and $L_i$ is the size of the box along the $i$ 
direction \cite{horowitz:08, schneider:13, schneider:14, schneider:16, 
nandi:18}. 
Recall that in this work we consider cubic boxes and, thus, $L_x=L_y=L_z=L$.

\subsection{Domains and defects}\label{ssec:domains}

All systems simulated for this work have a constant number density $n = 
0.05\unit{fm}^{-3}$, constant temperature $T=1\unit{MeV}$, and proton fractions 
of either $Y_p=0.30$ or $Y_p=0.40$. 
As discussed in Fig. 1 of Ref. \cite{berry:16} for a $Y_p=0.40$ system with 
40\,000 nucleons, protons and neutrons in the simulation volume initially bind 
locally due to the short-range nuclear attraction to form high density 
filaments. 
Over time, fluctuations due to the Coulomb repulsion introduce long-range 
correlations in the system and the filaments rearrange themselves. 
For systems with proton fraction $Y_p=0.30$ parallel plates perforated by an 
hexagonal arrangement of circular holes, the ``waffle'' phase 
\cite{schneider:14}, form. 
Meanwhile, for proton fraction $Y_p=0.40$ the system arranges itself in 
parallel plates connected by helical ramps, known ``parking garage'' 
structures \cite{berry:16}. 
The ``waffle'' phase \cite{schneider:14}, is similar to hexagonal networks seen 
in some phospholipid systems \cite{helfrich:73, seddon:95, marrink:04}, 
while the ``parking garage'' structure analog in lipid systems is known 
as Terasaki ramps \cite{terasaki:13, guven:14}.

The equilibration time of an MD system simulated at constant density, 
temperature, and proton fraction is correlated with the number of nucleons in 
the simulation volume, but depends in non-trivial ways on the proton fraction. 
As in previous works, we loosely define equilibrium as convergence of the mean 
and Gaussian curvatures of the system \cite{schneider:14, schneider:16}. 
Systems with a few thousand nucleons equilibrate rather quickly, while systems 
with hundreds of thousands to millions of nucleons take a significant amount of 
time to equilibrate. 
Some works suggest that different pasta phases may coexist \cite{okamoto:12, 
fattoyev:17, newton:18}, however, we do not observe that in any of our runs. 
Nonetheless, for volumes large enough we observed that the plates formed in our 
simulations could be oriented across multiple directions, \ie some runs 
exhibited more than one domain. 
To determine the formation time of domains and whether they were stable or 
eventually all merged into a single domain we implemented an algorithm to 
examine to which domain each proton in the simulation belonged to. 
This algorithm, discussed next, can be easily extended to include the neutrons. 
Since most neutrons are highly correlated with protons while a few form a low 
density background gas we do not include any neutrons in our analysis for the 
sake of speed.

The first step in our algorithm is to compute the proton \textit{elastic} 
structure factor $S^{e}_p(\boldq) = \langle \rho_i^\star(\boldq,t) 
\rho_i(\boldq,t) \rangle_t$, \ie the first right-hand-side term in Eq. 
\eqref{eq:sq}. 
The time average is performed over the last $10^6\unit{fm}/c$ of each run.
For the topologies studied in this work $S^{e}_p(\boldq)$ is much larger 
than the angle average 
$S^{e}_p(q) = \langle S^{e}_p(\boldq) \rangle_{\vert{q}\vert}$ whenever 
$\boldq$ is parallel to a direction normal to one of the plates formed. 
Mathematically, $S^{e}_p(\boldq) \gg \langle S^{e}_p(\boldq) 
\rangle_{q=\vert{\boldq}\vert} \Leftrightarrow \boldq \parallel 
\boldn_{\rm plates}$, where ${\boldn}_{\rm plates}$ is the direction 
normal to the plates 
(domains) in the system. 
If there is more than one domain, there will be multiple $\boldsymbol{n}_{i,\rm 
plates}$ and as many $\boldq_i$ that satisfy $S^{e}_p(\boldq_i) \gg S^{e}_p(q)$ 
where the $\boldq_i \parallel \boldsymbol{n}_{i,\rm plates}$. 
We note that the magnitude of $q = \vert\boldq\vert \sim 2\pi/d$, where $d$ is 
the average distance between nucleons in neighboring plates \cite{schneider:14, 
schneider:16}.

Once we have computed the set of momenta $\boldq_i = 2\pi (n'_x/L_x, 
n'_y/L_y, n'_z/L_z)$ such that $S^{e}_p(\boldq')\gg S^{e}_p(q)$, we compute a 
separate elastic structure factor for each proton $j$ in the system for each 
$\boldq_i$, \ie 
\begin{equation}\label{eq:seq}
\mathcal{S}_j(\boldq_i,t) = \rho^\star _j(\boldq_i,t) \rho_j(\boldq_i,t)
\end{equation}
where 
\begin{equation}\label{eq:rhoq}
 \rho_j(\boldq_i,t) = \frac{1}{\sqrt{\mathcal{N}_j(t)}} 
\sum_{k=1}^{\mathcal{N}_j(t)} e^{i\boldq_i \cdot 
(\boldsymbol{r}_j(t) - \boldsymbol{r}_k(t))}.
\end{equation}
Note that the subscripts $j$ in $\rho_j$ in Eqs. \eqref{eq:seg} and 
\eqref{eq:rhoq} are labels for each proton and not for nucleon type as in Sec. 
\ref{ssec:sq}. 
The sum in $k$ above only runs over the $\mathcal{N}_j(t)$ neighboring protons 
of $j$ at time $t$. 
The neighbors are defined as 
\begin{equation}
k \in \mathcal{N}_j(t) \Leftrightarrow  
\begin{cases}
 \left\vert x_j(t)-x_k(t)\right\vert \leq \vert L_x/2n'_x \vert,\\
 \left\vert y_j(t)-y_k(t)\right\vert \leq \vert L_y/2n'_y \vert,\\
 \left\vert z_j(t)-z_k(t)\right\vert \leq \vert L_z/2n'_z \vert.
 \end{cases}
\end{equation}
where $r_j(t)=(x_j(t),y_j(t),z_j(t))$ and similar for the index $k$. 
In cases where one or two of the $n'_w=0$, where $w=x$, $y$, or 
$z$, we set $n'_w\rightarrow10$ in the computations of the neighbor list only. 
This choice does not significantly affect $\mathcal{S}_j(\boldq_i,t)$ since, 
$n'_w=0$ if and only if there are no long range correlations along the $w$ 
direction(s).

After computing $\mathcal{S}_j(\boldq_i,t)$ we assign a proton $j$ to domain 
$D_i$ for which $\mathcal{S}_j(\boldq_i,t)$ is a maximum; 
unless it falls below a threshold, in which case it is set to the defects 
domain $D_0$. 

\begin{figure}[!htb]
\centering
\includegraphics[width=0.45\textwidth]{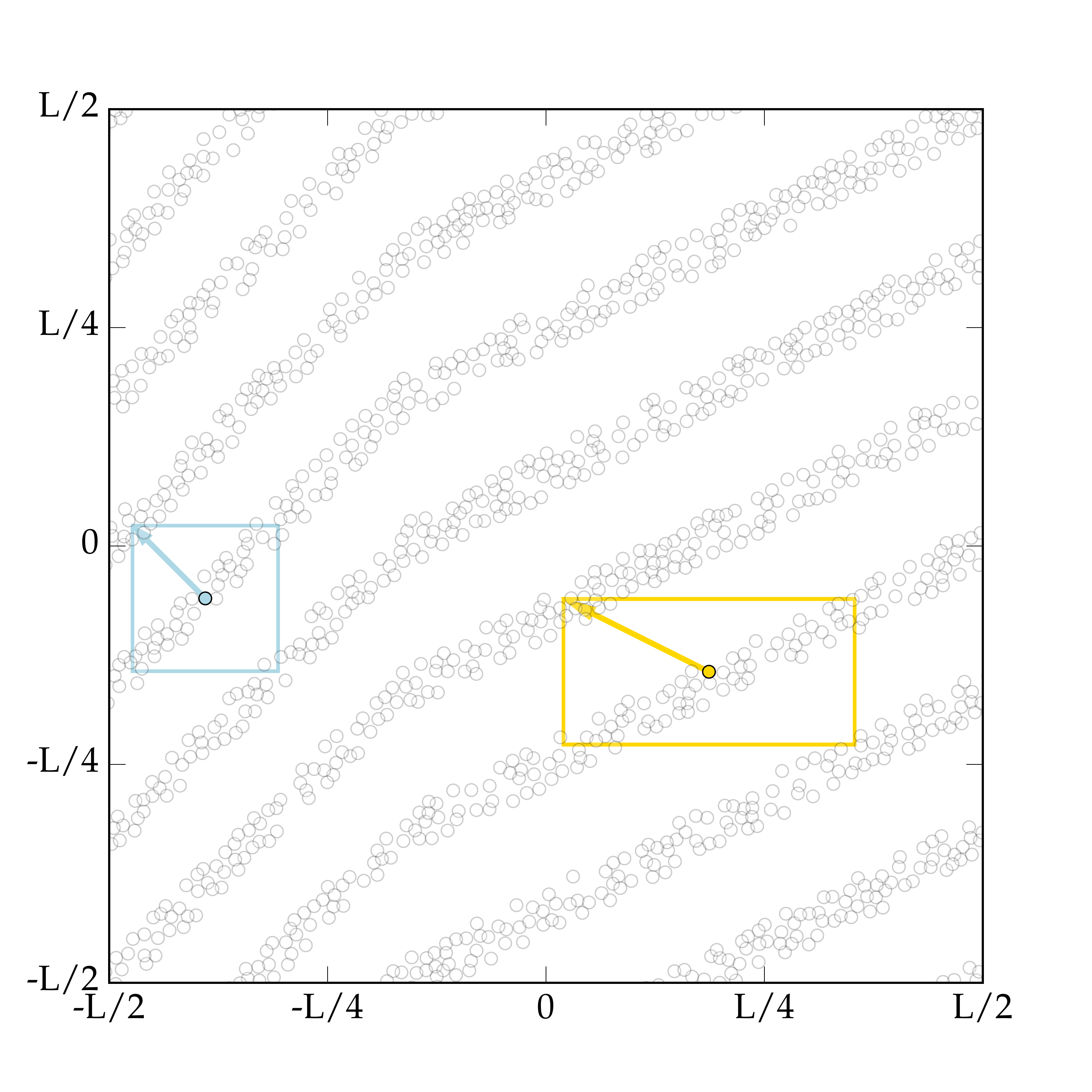}\\
\includegraphics[width=0.45\textwidth]{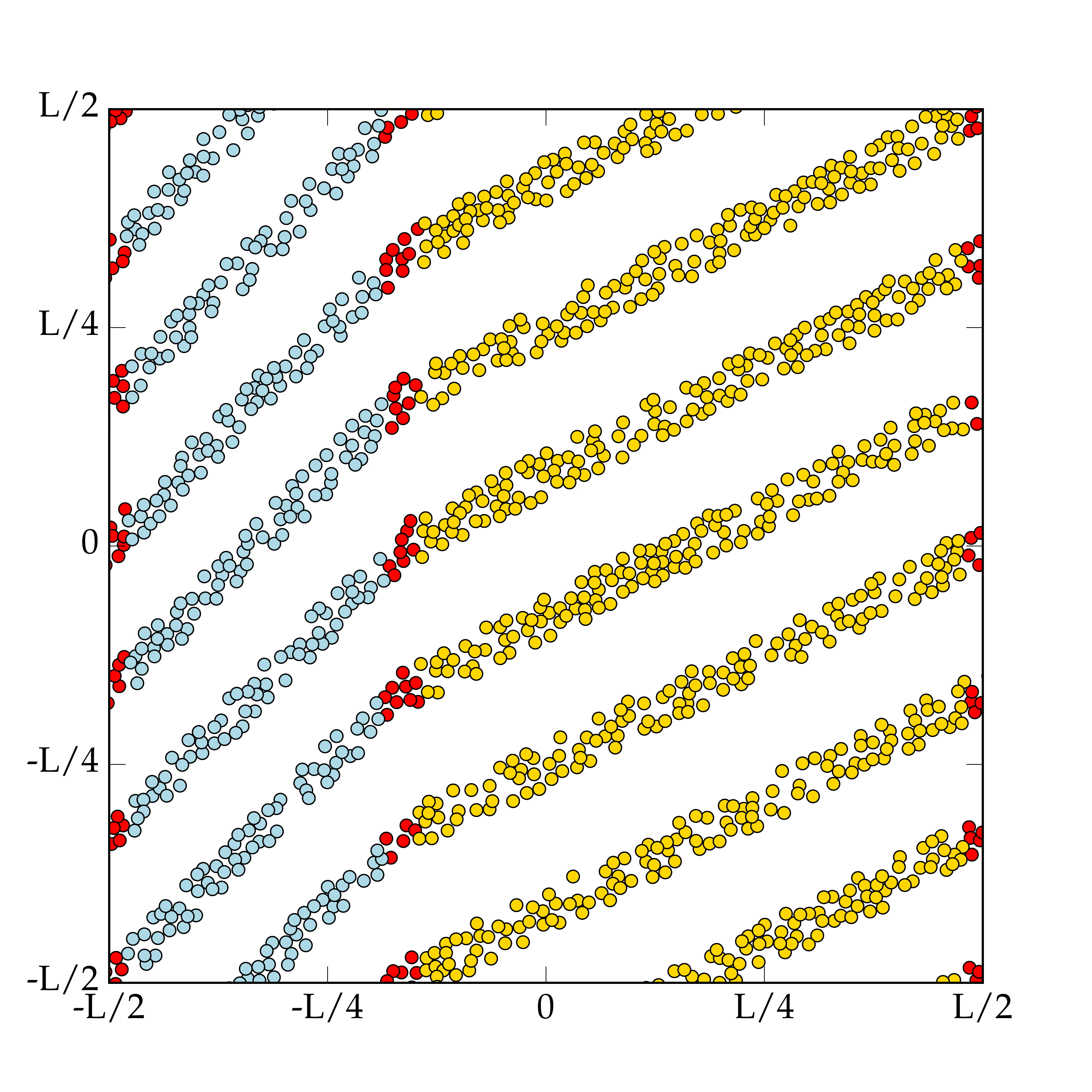}
\caption{\label{fig:2D} (Color online) Two dimensional example of our domain 
recognition algorithm. On the top we identify two particles that belong to 
each of the two domains formed, see text. On the bottom we color each particle 
according to the domain they belong to: red for defects domain $D_0$, yellow 
for domain $D_1$, identified by the vector $\boldq_1=\tfrac{2\pi}{L}(-3,6)$, 
and light blue for domain $D_2$, identified by the vector 
$\boldq_2=\tfrac{2\pi}{L}(-6,6)$.}
\end{figure}

A two dimensional example of our algorithm is shown in Fig \ref{fig:2D}. 
The system has periodic boundary conditions and its particles arranged 
themselves into two separate domains. 
For region $R_1$, defined by $-L/6\leq x\leq +L/2$, most particles form 
planes normal to the vector $\boldn_1 = (-L/3,L/6)$. 
For particles $r$ belonging to those planes the momentum transfer that 
maximizes $S_r(q)$ (here the time variable $t$ is omitted for clarity) is 
$\boldq_1 = \tfrac{2\pi}{L}(-3,6)$. 
One such example is shown by the particle tagged in yellow in the top panel of 
Fig. \ref{fig:2D}. 
Its $\mathcal{N}_r$ neighbors are the ones inside the yellow box, 
which can be regarded approximately as a unit cell for the planes in $R_1$. 
Thus, for most particles in $R_1$ $\mathcal{S}_r(\boldq_{1}) \gg 
\mathcal{S}_r(\boldq_{i\ne1})$ and we set them as being part of domain $D_1$.

Performing a similar analysis for the particles $l$ in the region 
$R_2$ defined by $-L/2\leq x \leq -L/6$ we obtain that the planes formed 
are normal to the vector $\boldn_2 = (-L/6,L/6)$. 
Thus, for particles in $R_2$ we obtain that $\mathcal{S}_l(\boldq_{2}) \gg 
\mathcal{S}_l(\boldq_{i\ne2})$ if and only if we set $\boldq_2 = 
\tfrac{2\pi}{L}(-6,6)$. 
These particles form domain $D_2$. 
One such particle is tagged in light-blue in the top panel of Fig. \ref{fig:2D} 
and its $\mathcal{N}_l$ neighbors are the particles inside the light-blue box.

Exceptions happen for particles $t$ near transition regions between different 
domains. 
For those particles, both $\mathcal{S}_t(\boldq_i)$ may have similar values. 
We identify the particle as belonging to the domain $D_i$ that produces the 
larger $\mathcal{S}_t(\boldq_i)$, unless this maxima is below a threshold
value. 
The threshold value is adjusted so that at the end of each run the number of 
particles that are on the ``defects'' domain $D_0$ is small while at the 
same time guarantees that the domains $D_i$ are clearly identified.

The method described above proved very accurate to identify different domains in 
our simulations. 
Its main limitation is that, due to thermal fluctuations of the domains,  
the angle $\theta_{ij}$ between the normal that defines two domains $i$ and $j$ 
has to be such that $\theta_{ij} \gtrsim 5 \degree.$ 
If that constraint is not imposed, often particles in domain $D_i$ ($D_{j}$) 
are misidentified as being part of $D_{j}$ ($D_i$).

In the bottom panel of Fig. \ref{fig:2D} we color all particles according to 
the domain they belong following our algorithm. 
Two domains are clearly identified with particles that form their 
interface being identified as ``defects''.

Although we only discuss cubic volumes in this work, the algorithm was presented 
in the more general framework of cuboids since it is used in the companion 
paper, Ref. \cite{caplan:18}, to study the breaking mechanism of nuclear pasta 
under extreme deformations.

\section{Results}\label{sec:Results}

We discuss the results for our molecular dynamics (MD) simulations with proton 
fractions $Y_p=0.30$, Sec. \ref{ssec:yp30}, and $Y_p=0.40$, Sec. 
\ref{ssec:yp40}.

\subsection{Simulations with $Y_p=0.30$}\label{ssec:yp30}

We start examining five runs with proton fraction $Y_p=0.30$. 
Two of these runs, the ones containing 51\,200 and 409\,600 nucleons, have 
already been presented under a different light in Ref. \cite{schneider:14}. 
We add to those two, three larger simulations containing 819\,200, 
1\,638\,400, and 3\,276\,800 nucleons. 
A summary of the runs is shown in Table \ref{Tab:yp30}.

\begin{table}[htb!]
\caption{\label{Tab:yp30} Summary of our MD runs with $Y_p=0.30$.
We list the number of nucleons in the first column, the side length of the 
simulation cube on the second column, the total evolution time in the third 
column, and the number of domains observed at the end of the run in 
the fourth column. In the last column $+1$ denotes that there is still a 
``defects'' domain at the end of the run, see text and Figs. \ref{fig:0819200}, 
\ref{fig:1638400} and \ref{fig:3276800}.}
\begin{ruledtabular}
\begin{tabular}{r c c r}
Nucleons & $t_{\mathrm total}$ & $L_{\mathrm box}$ & Domains \\
         & ($10^6$ fm/$c$)     &     (fm)          &   \\
\hline
      51\,200  &   31.0        &  100.8 &    1    \\
     409\,600  &   32.5        &  201.6 &    1    \\
     819\,200  &   55.0        &  254.0 &  1+1    \\
  1\,638\,400  &   37.0        &  320.0 &    1    \\
  3\,276\,800  &   32.0        &  403.2 &  5+1    \\
\end{tabular}
\end{ruledtabular}
\end{table}

We reiterate that all of our runs are performed within a cubic volume with 
constant nucleon number density $n = 0.05\unit{fm}^{-3}$, temperature 
$T=1\unit{MeV}$, and fixed screening length $\lambda=10\unit{fm}$.
Under these conditions all simulations with proton fraction $Y_p=0.30$ 
converged to the expected ``waffle'' phase \cite{schneider:14}. 
This same phase has been obtained by Sébille \etaln, albeit at a different 
proton fraction, by solving the equations of motion of single particle wave 
functions spanned in a wavelet basis where nucleons interact via a zero-range 
effective interaction \cite{sebille:11}. 
Sagert \etal also see the waffle phase from self-consistent Skyrme Hartree-Fock 
(SHF) calculations \cite{sagert:16}. 
However, in Sagert \etal the initial conditions for their SHF computation was 
obtained from the final configuration of an MD run. 
Thus, it is unclear if the final configuration in their simulations is a stable 
or meta-stable state.

\begin{figure}[!htb]
\centering
\includegraphics[width=0.45\textwidth]{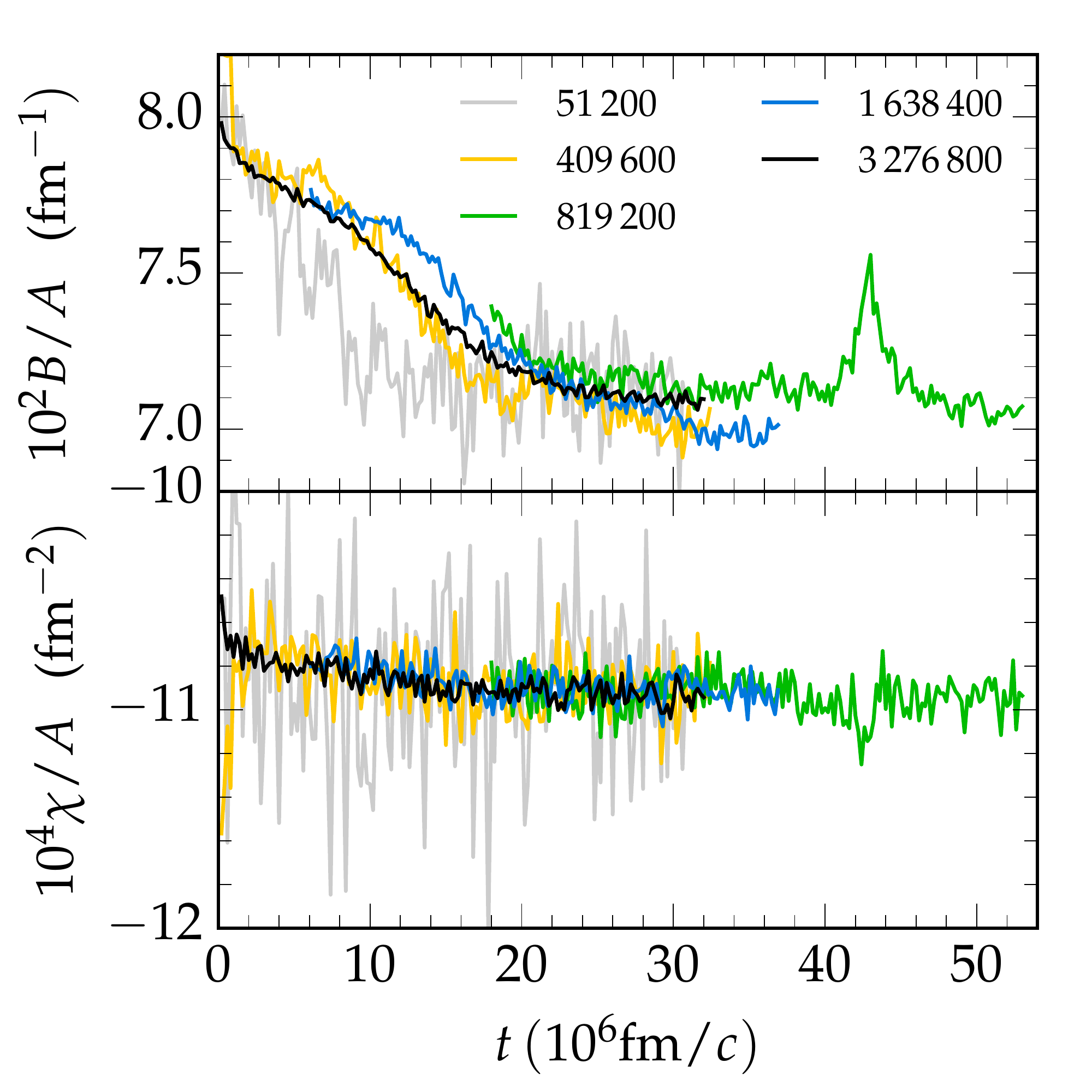}
\caption{\label{fig:M_30} (Color online) Plots of the normalized mean curvature 
$B/A$ (top) and normalized mean Gaussian curvature $\chi/A$ (bottom) as a 
function of simulation time $t$ for four simulations with $Y_p=0.30$, 
$n=0.05\unit{fm}^{-3}$ and $T=1.0\unit{MeV}$.}
\end{figure}

The topology of nuclear pasta is often characterized by its Minkowski 
functionals, see Refs. \cite{watanabe:03, watanabe:04, dorso:12, 
schuetrumpf:13, schneider:13, schuetrumpf:14, schneider:14}.
Specifically, the mean and Gaussian curvatures tell us about the degree of 
connectivity of the structures formed \cite{dorso:12, schuetrumpf:13}. 
In Fig. \ref{fig:M_30} we show the mean curvature $B$ and Gaussian curvature 
$\chi$ normalized by the surface area $A$ of the system for the $Y_p=0.30$ 
simulations. 
Technical details on how we compute Minkowski functionals are discussed in 
Refs. \cite{schneider:13, schneider:14}. 
The curvatures of all our simulations follow a similar pattern and results 
for the three new large simulations agree qualitatively with the two smaller 
ones\footnote{Due to a system purge of the Titan supercomputer files and 
incomplete backup of our data configurations for the 819\,200 nucleon run before 
$18 \cdot 10^6\unit{fm}/c$ and 1\,638\,400 before $6\cdot10^6\unit{fm}/c$ were 
lost and, thus, not plotted.}.
However, it is unclear whether any quantitative differences in the curvature 
are due to finite size effects or the presence of defects and/or multiple 
domains in the simulation. 
Furthermore, the 819\,200 simulations seems to go through a phase rearrangement 
between $40$ and $44\cdot10^6\unit{fm}/c$ where the curvatures deviate from 
their average values. 
This deviation is similar to that what is observed for the bond angle metric 
$Q_6$ and the diffusion coefficient of ions in Coulomb crystals as it freezes 
\cite{hughto:11}.

\begin{figure}[!htb]
\centering
\includegraphics[width=0.45\textwidth]{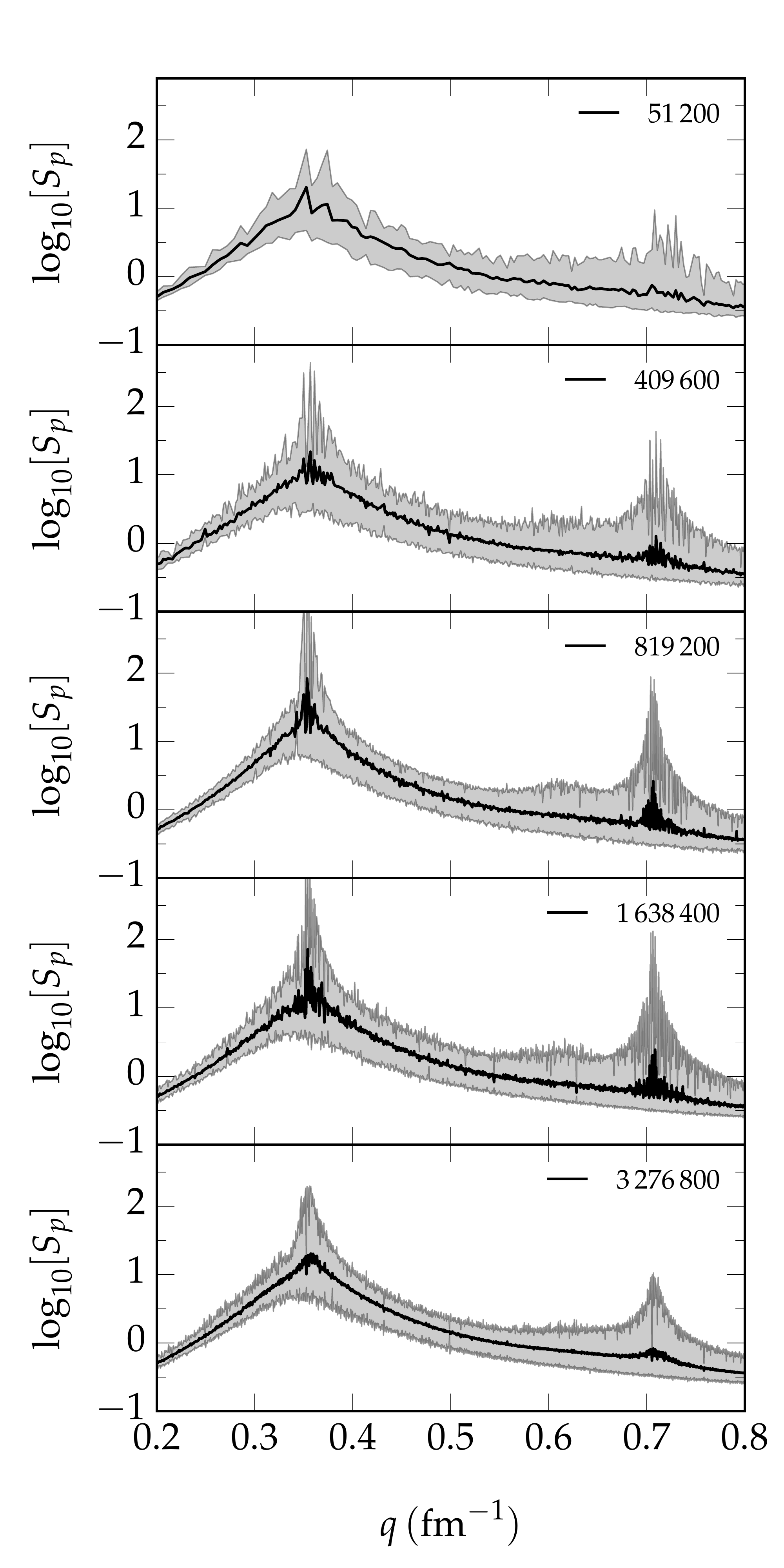}
\caption{\label{fig:sq_30} Plots of the angle averaged proton structure factor 
$S_p(q)=\langle{S_p(\boldq)}\rangle_{q}$ (thick black lines) for the 
final $1.0\cdot10^6\unit{fm}/c$ of each simulation. The average value is 
bound by the maximum and minimum in $S_p(q)$ for each $q=\vert\boldq\vert$ 
(shaded grey area).}
\end{figure}

Besides the average curvatures, another important quantity to measure from 
these large simulations is the structure factor of each topology as they 
encode the transport properties of the pasta phases \cite{nandkumar:84, 
horowitz:04a, chugunov:05, horowitz:08, dorso:12, horowitz:15, yakovlev:15, 
schneider:16, nandi:18}. 
We follow our previous work \cite{schneider:16} and compute the proton 
structure factors $S_p(\boldq)$ for possible values of $\boldq$ within our 
periodic simulation box, Eq. \eqref{eq:q}. 
If we assume that pasta has multiple uncorrelated domains it is convenient to 
describe its structure factor as an average over all momentum transfers with 
same magnitude $q=\vert\boldq\vert$, \ie obtain $S_p(q) = \langle 
{S_p(\boldq)} \rangle_{q}$ \cite{horowitz:04a, schneider:13, nandi:18}. 
It is also possible that domains with different orientations are only stable 
when separated by distances larger than the size of our simulation volume and,
thus, even though our simulations may only show a single domain, the relevant 
quantity is still the angle averaged quantity $S_p(q)$. 
However, it may be that the pasta phases are in fact anisotropic over very 
large length scales or that its defects are correlated \cite{horowitz:15, 
schneider:16} and, thus, the anisotropy in $S_p(\boldq)$ does affect its 
transport properties. 
In Fig. \ref{fig:sq_30} we show the angular average structure factor for 
protons $S_p(q)$ and its upper and lower bounds, defined by the maxima and 
the minima in $S_p(\boldq)$ for a given $q=\vert\boldq\vert$.

By comparing the results of simulations of different sizes, it is clear that 
the 51\,200 nucleon run is too small to reproduce some of the quantitative 
features in $S_p(\boldq)$ seen in the larger runs\footnote{An error in the 
$S_p(\boldq)$ code used in Ref. \cite{schneider:14} was corrected. Although 
results in Ref. \cite{schneider:14} are qualitatively correct, the error 
changes the magnitude of some of the peaks in $S_p(q)$ discussed in that work 
and, thus, direct comparison of those results and the ones presented here is 
not possible.}. 
The most obvious differences are the lower number and smaller magnitude of 
peaks in the 51\,200 nucleon run.
This may be due to a couple of factors such as the finite size of the 
simulation or the formation of multiple mostly independent domains in the 
simulation volume. 
However, we have shown in Fig. 3 of Ref. \cite{schneider:14} that the 51\,200 
nucleon simulation forms only a single domain. 
Furthermore, we will show below that three of the four simulations with more 
than 51\,200 nucleons form a single dominant domain at the end of the run. 
Therefore, the culprit of the differences seen between $S_p(\boldq)$ for 
the 51\,200 simulation and the larger ones is the simulation size. 
If this is the case, it introduces a severe constraint in the computations of 
transport properties of nuclear pasta.
Even though MD simulations containing 51\,200 nucleons can now be easily 
achieved with the IUMD code, quantum molecular dynamics (QMD) simulations, 
which use more sophisticated interaction potentials between nucleons 
\cite{watanabe:09, nandi:18}, as well as quantum-mechanical state-of-the-art 
pasta simulations \cite{schuetrumpf:14, sagert:16, fattoyev:17}, which rely on 
energy density functional calculations, are still limited, to a few dozen 
thousand nucleons or much less. 
For example, recently, Nandi and Schramm computed structure factors and Coulomb 
logarithms from QMD simulations for a range of densities, temperatures, and 
proton fractions \cite{nandi:18}. 
All of their simulations contain 8\,192 to 16\,384 nucleons. 
Assuming our results also hold for simulations that use different 
nucleon-nucleon interactions, it is likely that the results for transport 
properties of Nandi and Schramm still suffer from considerable finite size 
effects. 
Nevertheless, it is encouraging that there is a very good qualitative 
agreement between their results for the structure factor $S_p(q)$ and ours for 
$Y_p=0.30$, $n \sim 0.5\unit{fm}^{-3}$ at $T=1\unit{MeV}$.

As we increase the number of nucleons from 51\,200 nucleons to 409\,600 the 
box length along each direction doubles and, therefore, the number of vectors 
$\boldq$ to be analyzed as well as the statistical significance of our results 
increase by a factor of $\sqrt{8}$. 
The magnitudes of the peaks in $S_p(\boldq)$ as well as the number of 
oscillations in both $S_p(\boldq)$ and $S_p(q)$ near $q' \sim 0.36 
\unit{fm}^{-1}$ and $q'' \sim 2q'$ increase considerably with a larger 
simulation\footnote{The magnitude $q'\sim2\pi/d$ is directly related to the 
average distance $d$ between nucleons in neighboring plates in the simulation 
volume.}.

By increasing further the simulation volume, to 819\,200 nucleons, the maxima 
in $S_p(\boldq)$ and its average $S_p(q)$ increase even more in magnitude 
near $q'$ and $q''$. 
However, there is little quantitative difference between $S_p(\boldq)$ and 
$S_p(q)$ between the runs with to 819\,200 and 1\,638\,400 nucleons.

In our largest run, with 3\,276\,800 nucleons, the structure factor 
$S_p(\boldq)$ is qualitatively very similar to the ones computed for the 
smaller simulations. 
The peaks in $S_p(\boldq)$ for this run, however, are somewhat smaller than 
the ones for the 819\,200 and 1\,638\,400 nucleon runs. 
We show below that this is likely not due to finite size effects, which should 
be well constrained in a simulation of this size, but due to this simulation 
having multiple large domains in the time we analyzed its structure factor.
This is unlike the smaller simulations, which by the end of the run show a 
single large domain, which occupies almost all of the simulation volume.

Despite the seemingly convergence of the curvatures, an interesting question 
to ask is whether the simulated systems, once evolved for a long time, are 
arranged into a single domain or multiple ones. 
We use the methods of Secs. \ref{ssec:sq} and \ref{ssec:domains} to identify
the main domain(s) in each run. 
These are discussed in detail for the three new simulations ran for this work.

\subsubsection{Simulation with  819\,200 nucleons}

From all of our simulations, the one with 819\,200 nucleons and proton 
fraction $Y_p=0.30$ was the one evolved for the longest time, about 
$55\cdot10^6\unit{fm}/c$. 
This run cost approximately $2.5\cdot10^5$ node\,hours on the hybrid CPU/GPU 
nodes of the Titan supercomputer.

\begin{figure}[!htb]
\centering
\includegraphics[trim=0 10 0 10, clip, width=0.45\textwidth] 
{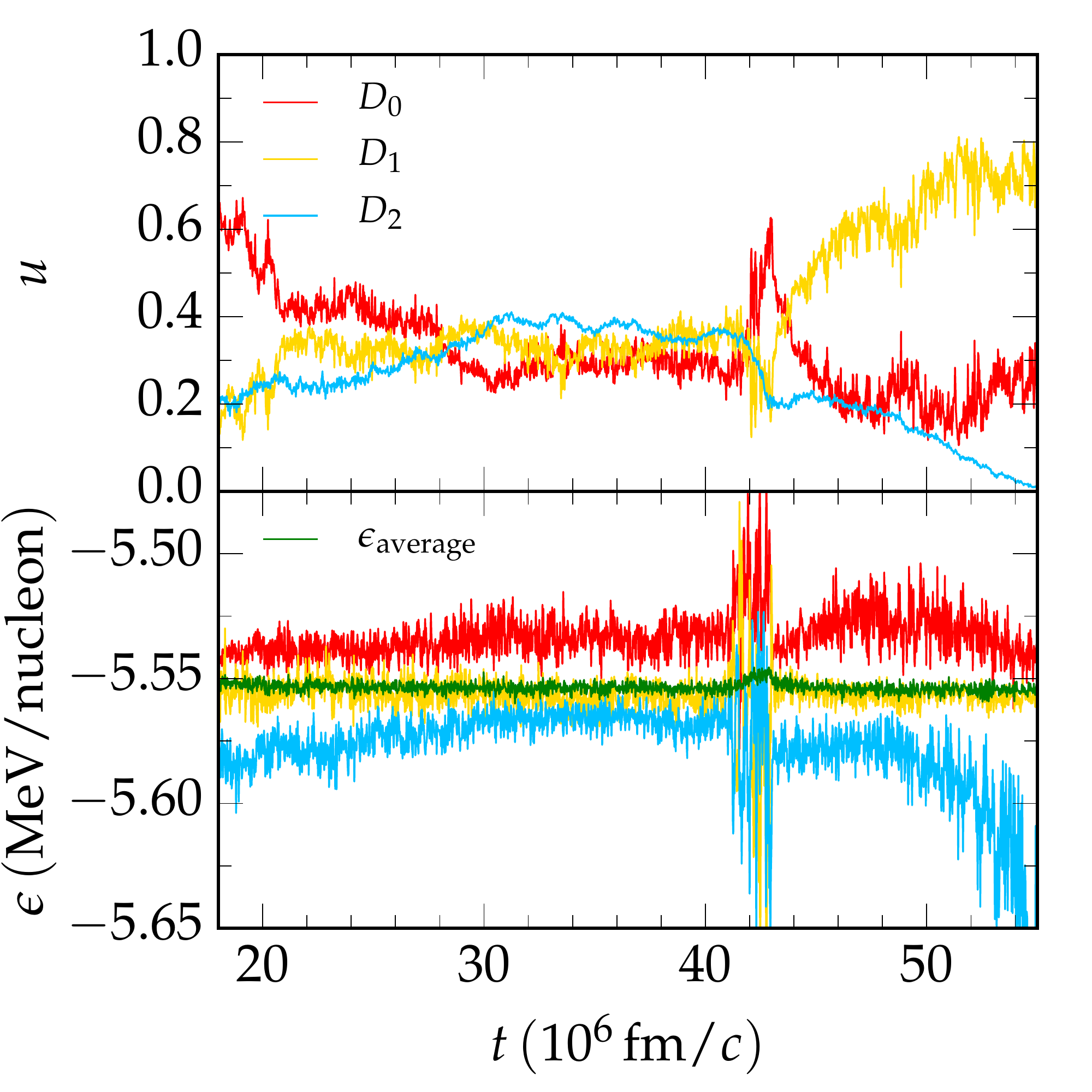}
\caption{\label{fig:e_0819200} (Color online) Volume fraction $u$ (top) and 
potential energy per nucleon $\epsilon$ (bottom) for each domain in the system 
for the 819\,200 nucleon simulation as a function of simulation time. 
The three domains are $D_0$ (defects), $D_1$ 
($\boldq_1=\tfrac{2\pi}{L}(-1,2,14)$), and $D_2$ 
($\boldq_2=\tfrac{2\pi}{L}(-8,9,8)$). 
Green line in the bottom pane is the average system energy. 
We note that due to a system purge of Titan files and incomplete backup of our 
data configurations for the 819\,200 nucleon run before $18 \cdot 
10^6\unit{fm}/c$ were lost.}
\end{figure}

In Fig. \ref{fig:e_0819200} we plot the volume fraction $u$ and energy per 
nucleon $\epsilon$ of two domains identified in the system in addition to a 
``defects'' domain.
Domains $D_1$ and $D_2$ are defined, respectively by the momenta
$\boldq_1=\tfrac{2\pi}{L}(-1,2,14)$ and $\boldq_2=\tfrac{2\pi}{L}(-8,9,8)$ 
where $L=240\unit{fm}$. 
We also define domain $D_0$ as the group of nucleons that are not part of either
$D_1$ nor $D_2$. 
Domain $D_0$ is usually formed by many small domains and/or the interface 
between domains $D_1$ and $D_2$.

At the start of the simulation the perforated plates formed do not have any 
particular orientation, and, thus, $D_0$ occupies almost 
all of the simulation volume (not shown). 
However, at $18\cdot10^6\unit{fm}/c$ the system has formed two main domains, 
each occupying about 20\% of the simulation volume. 
The domains are parallel plates with an hexagonal lattice of almost circular 
holes, the ``waffle'' phase discussed in Ref. \cite{schneider:14}. 
All three domains have similar volumes from $20$ to $40\cdot10^6\unit{fm}/c$, 
with neither dominating significantly over the other two. 
Furthermore, during this time there also little change in the average 
curvatures of the system, Fig. \ref{fig:M_30}. 
However, the average energy per nucleon $\epsilon_i$ of each domain $i$ follows 
a clear order, $\epsilon_2 < \epsilon_1 < \epsilon_0$. 
Although domain $D_2$ has a lower energy per nucleon than domain $D_1$, as 
domain $D_2$ increases in volume its energy per nucleon $\epsilon_2$ 
also increase, becoming similar to that of domain $D_1$. 
It is likely that if domain $D_2$ increased further in volume its average 
energy would become larger than that of domain $D_1$ and, thus, its 
growth is disfavored. 
Between $40$ and $44\cdot10^6\unit{fm}/c$ thermal fluctuations in the system 
force it to rearrange itself quickly. 
This is seen by abrupt changes in the volume and energy per nucleon of the 
three domains tracked. 
When this happens, domain $D_2$ decreases in volume until it almost disappears 
by the end of the run, $t=55\cdot10^6\unit{fm}/c$. 
In the final configuration, domain $D_1$ occupies 70\% of the simulation volume 
while domain $D_0$ (defects) occupy the remainder. 
It is likely that if this system is evolved for a longer time domain $D_1$ will 
occupy all of the simulation volume as is the case in smaller systems 
\cite{schneider:16}.

\begin{figure}[!htb]
\centering
\includegraphics[trim=0 30 20 0, clip, width=0.45\textwidth] 
{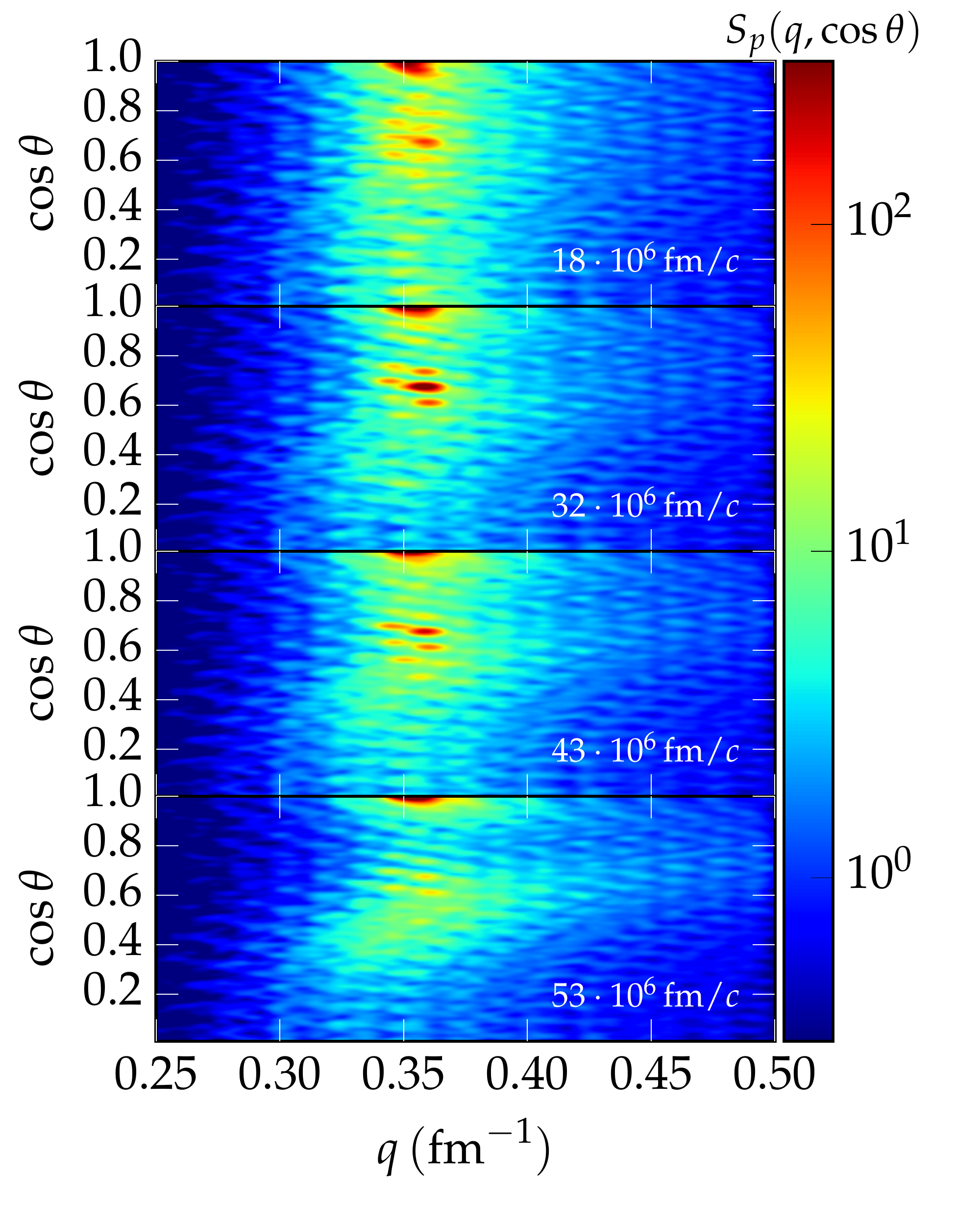}
\caption{\label{fig:sq_0819200} (Color online) Proton structure factor 
$S_p(q,\cos\theta)$ for the 819\,200 nucleon simulation as a function of the 
momentum transfer $q$ and the angle $\theta$ between $\boldq$ and the direction 
where $S_p^e(\boldq)$ is maximum, $\boldq_{\rm max}=\boldq_1$. 
This plot was generated from smoothing a 2D histogram of $S_p(q,\cos\theta)$ 
using a Gaussian filter with standard deviations $\sigma_q=0.025\unit{fm}^{-1}$ 
and $\sigma_{\cos\theta}=0.05$.}
\end{figure}

\begin{figure*}[!htb]
\centering
\includegraphics[width=0.8\textwidth]{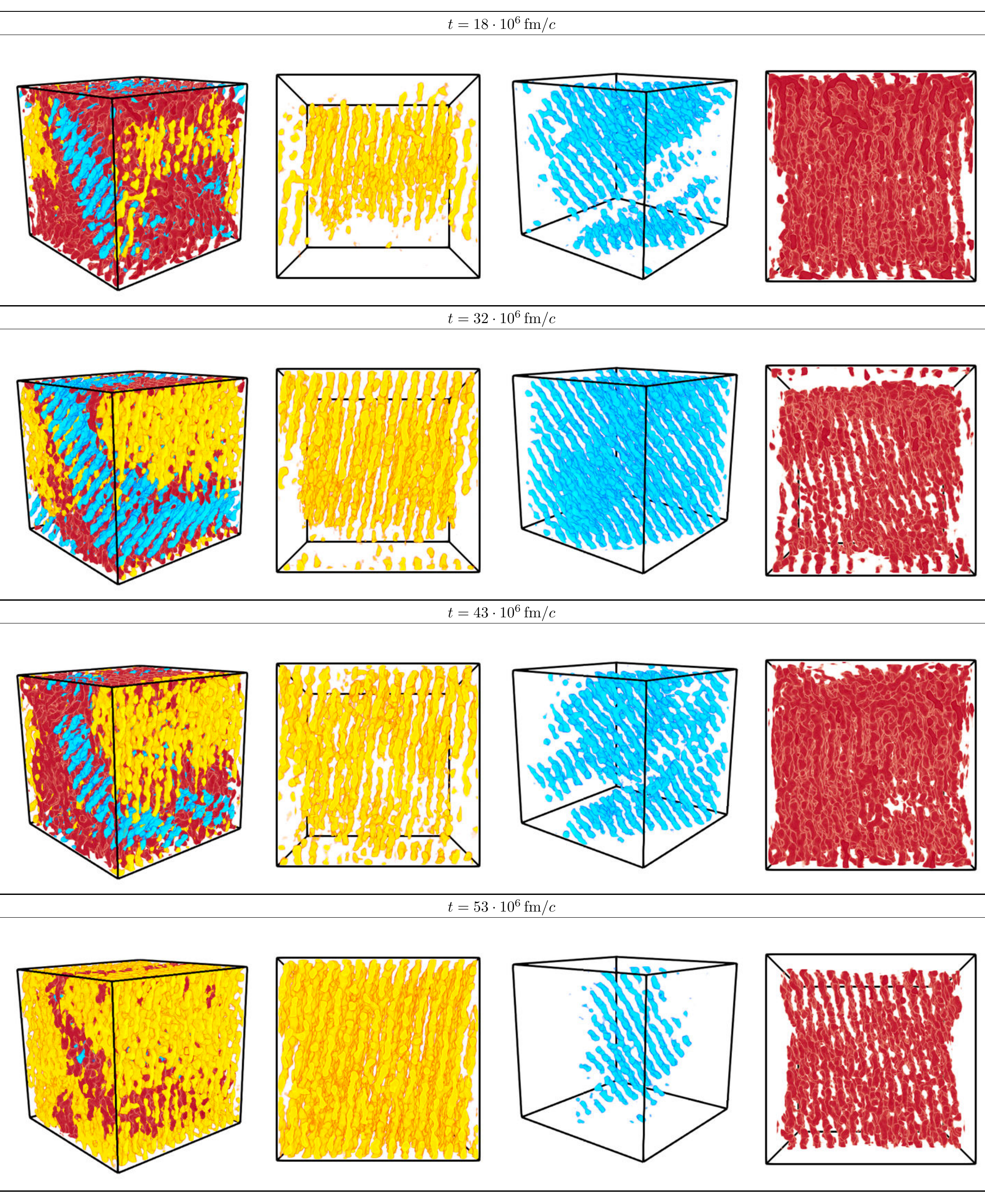}
\caption{\label{fig:0819200} (Color online) Configurations of our 
819\,200 nucleon run at four different times, $t=18,\,32\,43$ and 
$53\cdot10^6\unit{fm}/c$.
In the first column we show all domains: $D_0$ in red, $D_1$ in yellow, 
and $D_2$ in light blue. 
In columns 2, 3, and 4 we show, respectively, domains $D_1$, $D_2$ and $D_0$.}
\end{figure*}

In Figure \ref{fig:sq_0819200} we plot the proton structure factor averaged 
over the azimuthal angle, $S_p(q,\cos\theta)$, at four different times in our 
simulation. 
For a clearer image we smooth the 2D histogram of $S_p(q,\cos\theta)$ using a 
Gaussian filter with standard deviations $\sigma_q=0.025\unit{fm}^{-1}$ and 
$\sigma_{\cos\theta}=0.05$. 
We limit the plot to regions near $q\sim0.35\unit{fm}^{-1}$ which is where the 
first peak in the angle average $S_p(q)$ occurs, see Fig. \ref{fig:sq_30}. 
The angle $\theta(\boldq)$ is chosen such that $\theta=0$ ($\cos\theta=1$) is 
parallel to the direction $\boldq_{\rm max}$ where $S_p^e(\boldq_{\rm max}) = 
\max(S_p^e(\boldq))$ in the last configuration of our simulation, \ie
\begin{equation}\label{eq:theta}
 \cos\theta(\boldq) = \frac{\boldq\cdot\boldq_{\rm max}} 
{\vert \boldq \vert \vert \boldq_{\rm max}\vert}.
\end{equation} 
We note that the direction of $\boldq_{\rm max}$ coincides with $\boldq_1 = 
\tfrac{2\pi}{L}(-1,2,14)$, the direction we chose to define domain $D_1$, the 
largest one at the end of our simulation.  
Although this may seem obvious it is not always the case as discussed in Ref. 
\cite{schneider:14} and for our 3\,276\,800 nucleons run discussed below.

The changes in domain sizes over time, seen in Fig. \ref{fig:e_0819200}, 
can be inferred to a degree from the evolution of $S(q,\cos\theta)$ shown in 
Fig. \ref{fig:sq_0819200}. 
At $18\cdot10^6\unit{fm}/c$ the system shows two prominent peaks in 
$S_p(q,\cos\theta)$: one at $\boldq_1=\tfrac{2\pi}{L}(-1,2,14)$, 
$q\sim0.35\unit{fm}^{-1}$ and $\cos\theta=1$, and another 
at $\boldq_2=\tfrac{2\pi}{L}(-8,9,8)$, $q\sim0.36\unit{fm}^{-1}$ and 
$\cos\theta=0.67$. 
This implies an angle $\theta_{12}\sim48\degree$ between $\boldq_1$ and 
$\boldq_2$. 
As mentioned above, we used these two $\boldq_i$ to define domains $D_1$ and 
$D_2$. 
At this early time we see several other smaller peaks in $S_p(q,\cos\theta)$ in 
the range $0.34\unit{fm}^{-1} \lesssim{q} \lesssim 0.37\unit{fm}^{-1}$ and $0 
\lesssim{\cos\theta} \lesssim 1$. 
Each peak corresponds to a direction perpendicular to a small domain, likely 
included in the defects domain $D_0$, while their magnitudes are correlated 
with the volume each of these small domains occupies.

In Figure \ref{fig:0819200} we show the configuration of the domains $D_1$, 
yellow plates, $D_2$, light blue plates, and $D_0$, red plates, at four 
different times in our simulation. 
At $18\cdot10^6\unit{fm}/c$ the system is still dominated by the 
many small and likely uncorrelated domains that form $D_0$, Figs. 
\ref{fig:e_0819200} and \ref{fig:0819200}. 
Between $18$ and $32\cdot10^6\unit{fm}/c$ both domains $D_1$ and $D_2$ increase 
in volume while $D_0$ decreases. 
This can be inferred by the darkening and sharpening of the peaks in 
$S_p(q,\cos\theta)$ near $\boldq_1$ and $\boldq_2$ at $32\cdot10^6\unit{fm}/c$, 
Fig. \ref{fig:sq_0819200} and, even more clearly, in the second row of Fig. 
\ref{fig:0819200}. 
Moreover, the number and magnitude of peaks in $S_p(q,\cos\theta)$ for 
$\cos\theta\leq0.5$ decrease considerably when compared to the 
$18\cdot10^6\unit{fm}/c$ configuration, meaning that domains nearly 
perpendicular to the $D_1$ are disfavored. 
After $43\cdot10^6\unit{fm}/c$ in simulation time, domain $D_2$ decreases 
significantly in volume.
This is accompanied by a decrease in magnitude of $S_p(q,\cos\theta)$ near 
$\boldq_2$ and in volume of the light blue region, see Fig. \ref{fig:0819200}.
However, around that same time, small domains nearly perpendicular to the 
domain $D_1$ have formed, as seen by the reappearance of many small peaks in 
the region $q\sim0.35$ with $\cos\theta\lesssim0.5$.
Since we group these domains alongside others in $D_0$, domain $D_0$ increases 
in volume around that time, see Figs. \ref{fig:e_0819200} and  
\ref{fig:sq_0819200} and red region in third row of Fig. \ref{fig:0819200}. 
This change also correlates with a departure from average of the mean and 
Gaussian curvatures shown in Fig. \ref{fig:M_30}. 
Nearly the end of our run, $t=53\cdot10^6\unit{fm}/c$, domain $D_2$ has 
decreased to a very small volume which is separated from the domain $D_1$ by 
domain $D_0$. 
The near disappearance of domain $D_2$ and significant decrease in size 
of $D_0$ coincides with the disappearance, respectively, of the sharp peak in 
$S_p(q_2,\cos\theta_{12})$ and the decrease in the number and magnitude 
of peaks with $\cos\theta\lesssim0.5$ near $q\sim0.35$, bottom plot in Fig. 
\ref{fig:sq_0819200}.

\subsubsection{Simulation with 1\,638\,400 nucleons}

Our simulation run with 1\,638\,400 nucleons was equilibrated for 
$37\cdot10^6\unit{fm}/c$. 
This run cost approximately $3.2\cdot10^5$ node\,hours on the hybrid CPU/GPU 
nodes of the Titan supercomputer. 
Almost all of the nucleons arranged themselves in a single domain at the end of 
the run.

\begin{figure}[!htb]
\centering
\includegraphics[trim=0 10 0 10, clip, width=0.45\textwidth] 
{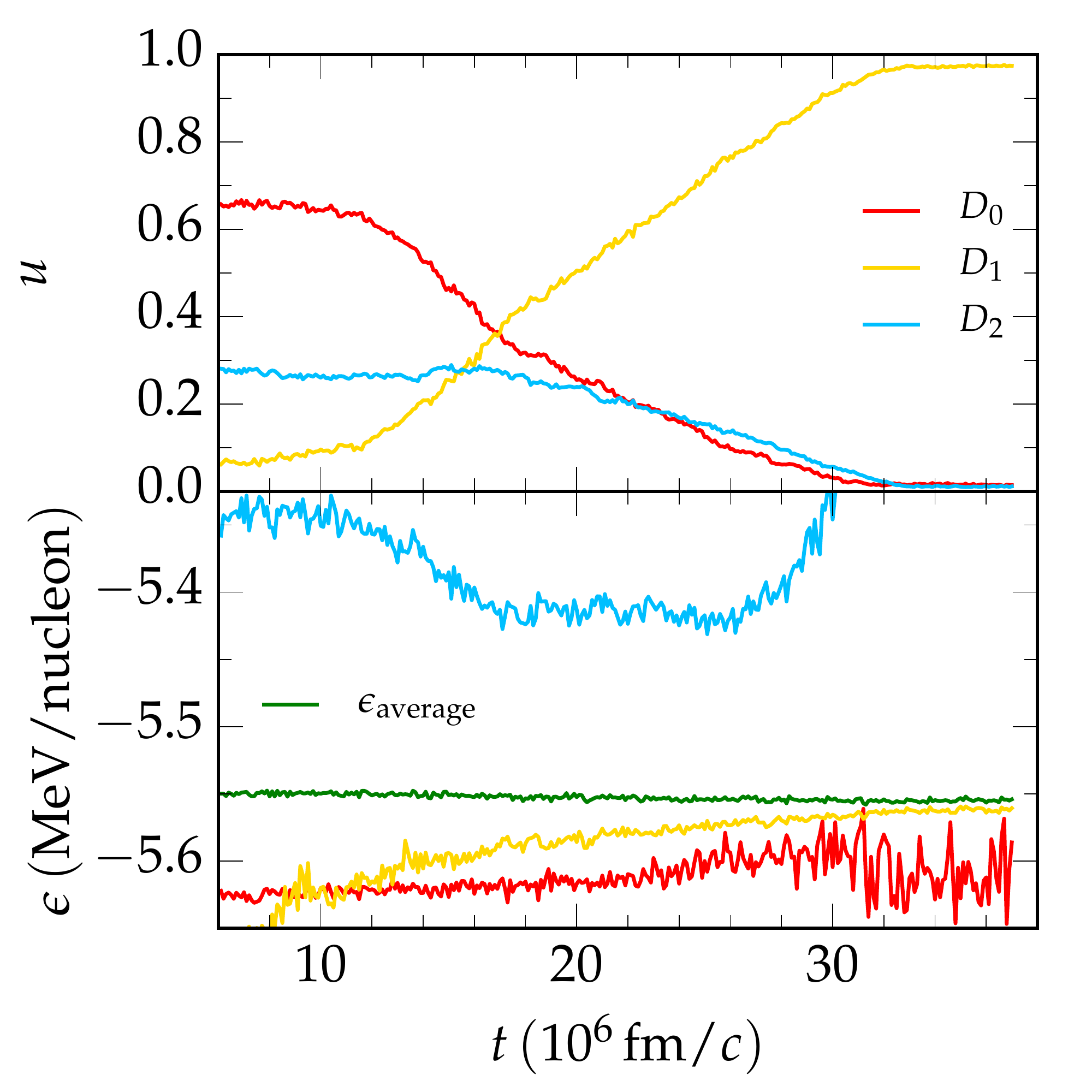}
\caption{\label{fig:e_1638400} (Color online) Volume fraction $u$ (top) and 
potential energy per nucleon $\epsilon$ (bottom) for each domain in the system 
for the 1\,638\,400 nucleon simulation as a function of simulation time. 
The three domains are $D_0$ (defects), $D_1$ 
($\boldq_1=\tfrac{2\pi}{L}(-11,11,9)$), and $D_2$ 
($\boldq_2=\tfrac{2\pi}{L}(-18,-1,1)$).
Green line in the bottom pane is the average system energy. 
We note that due to a system purge of Titan files and incomplete backup of our 
data configurations for the 819\,200 nucleon run before $6 \cdot 
10^6\unit{fm}/c$ were lost.}
\end{figure}

\begin{figure}[!htb]
\centering
\includegraphics[trim=20 0 20 0, clip, width=0.45\textwidth] 
{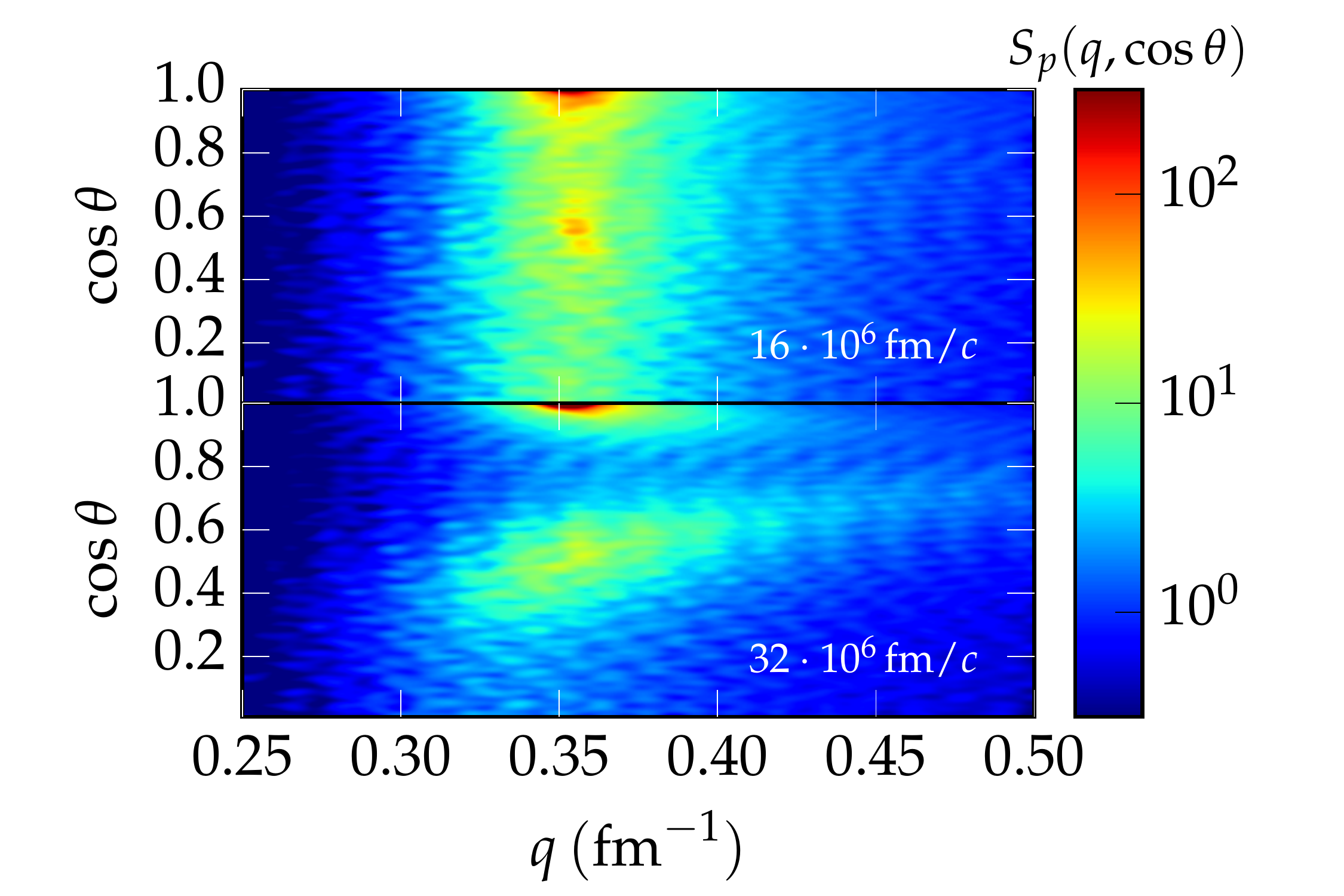}
\caption{\label{fig:sq_1638400} (Color online) Similar to Fig. 
\ref{fig:sq_0819200} but for the 1\,638\,400 nucleon system at times $t=16$ 
and $32\cdot10^6\unit{fm}/c$.}
\end{figure}

\begin{figure*}[!htb]
\centering
\includegraphics[width=0.8\textwidth]{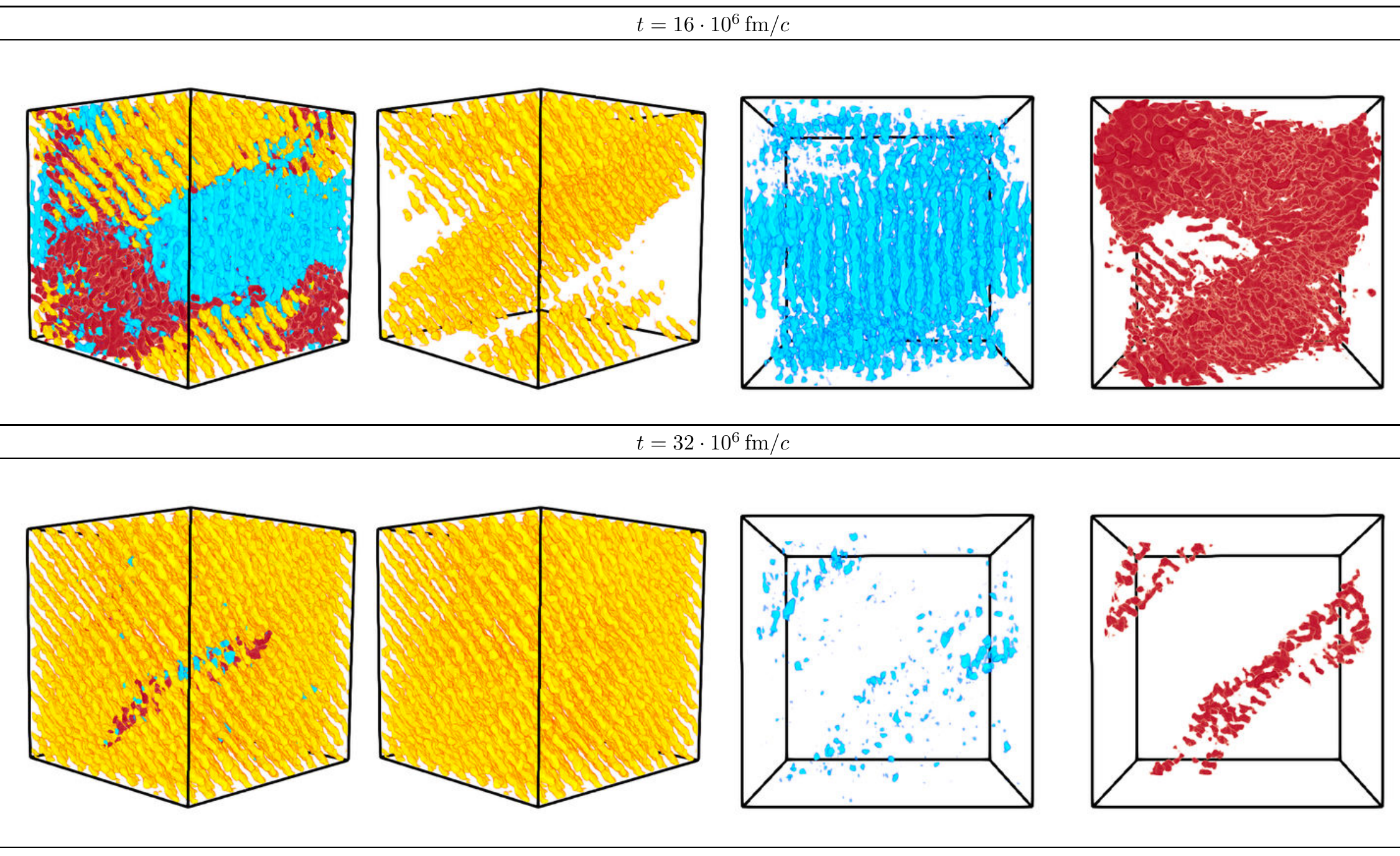}
\caption{\label{fig:1638400} (Color online) Similar to Fig. 
\ref{fig:0819200} but for the 1\,638\,400 nucleon system at times $t=16$ 
and $32\cdot10^6\unit{fm}/c$.}
\end{figure*}

We perform a data analysis like the one described for the $819\,200$ nucleon 
system. 
By computing $S(\boldq)$ halfway through the simulation we identify two 
dominant domains: $D_1$ defined by $\boldq_1=\tfrac{2\pi}{L}(-11,11,9)$ and 
$D_2$ defined by $\boldq_2=\tfrac{2\pi}{L}(-18,-1,1)$. 
Here $L=320\unit{fm}$. 
Similarly to the 819\,200 case, the angle between the two domains is 
$\sim53\degree$. 
Again we define $D_0$ as the set of nucleons that belong to neither $D_1$ or 
$D_2$.

From the data we have we observe that domain $D_2$ quickly grows in size and at 
$6\cdot10^6\unit{fm}/c$ already occupies 30\% of the simulation volume, top 
panel of Fig. \ref{fig:e_1638400}.
However, this domain has a significantly larger energy per nucleon than domain 
$D_1$, bottom panel of Fig. \ref{fig:e_1638400}. 
Thus, the latter is favored and quickly grows: by the end of the run both $D_0$ 
and $D_2$ have almost completely disappeared, while $D_1$ occupies almost all of 
the simulation volume.
This progression can also be inferred from the evolution of the peaks in 
$S_p(q,\cos\theta)$, plotted in Fig. \ref{fig:sq_1638400}, and explicitly shown
in Fig. \ref{fig:1638400}.

\subsubsection{Simulation with 3\,276\,800 nucleons}

The simulation with 3\,276\,800 nucleons is the largest one in our work and, to 
our knowledge, the largest nuclear pasta simulation performed to date. 
This run was performed exclusively on the hybrid CPU/GPU nodes of the Big Red 2 
supercomputer and cost approximately $1.9\cdot10^6$ node\,hours. 
Despite its long run time, this systems is still composed of several domains in 
its final configuration at $t=32\cdot10^6\unit{fm}/c$.

\begin{figure}[!htb]
\centering
\includegraphics[trim=0 10 0 10, clip, width=0.45\textwidth] 
{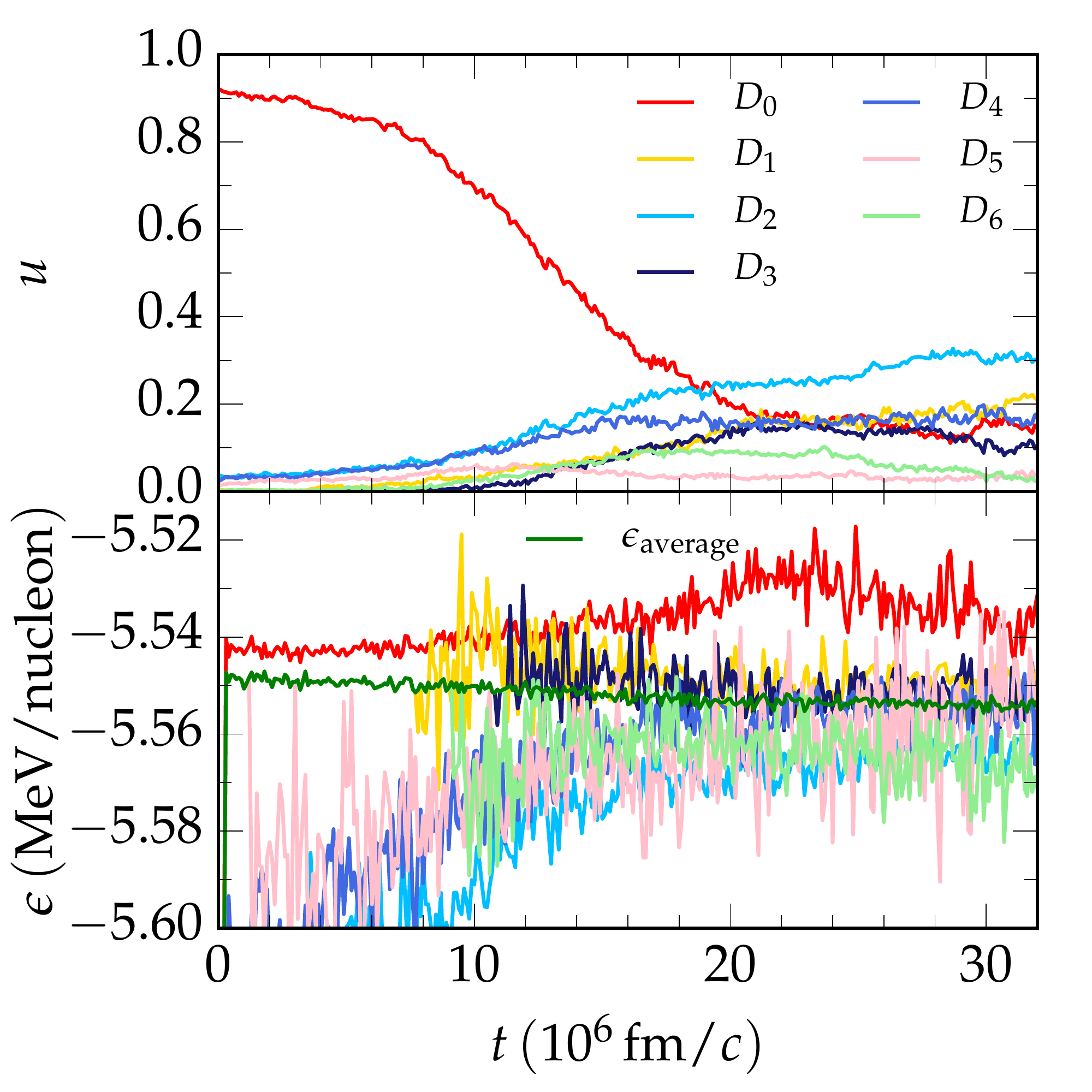}
\caption{\label{fig:e_3276800} Volume fraction $u$ (top) and 
potential energy per nucleon $\epsilon$ (bottom) for each domain in the system 
for the 3\,276\,800 nucleon simulation as a function of simulation time. 
The seven domains are 
$D_0$ (defects), 
$D_1$ ($\boldq_1=\tfrac{2\pi}{L}(-10,19, 7)$), 
$D_2$ ($\boldq_2=\tfrac{2\pi}{L}(-11,13,15)$), 
$D_3$ ($\boldq_3=\tfrac{2\pi}{L}( -4, 8,21)$), 
$D_4$ ($\boldq_4=\tfrac{2\pi}{L}(-14,15,10)$),
$D_5$ ($\boldq_5=\tfrac{2\pi}{L}(-16, 8,14)$), and
$D_6$ ($\boldq_6=\tfrac{2\pi}{L}(-13,18,-5)$).
Green line in the bottom pane is the average system energy. 
To reduce noise in the plot of $\epsilon$ we do not show the values for domains 
at times when their volume fraction is $u<0.02$.}
\end{figure}

In Fig. \ref{fig:e_3276800} we plot the volume fraction $u$ and potential 
energy per nucleon $\epsilon$ for seven domains. 
These domains are 
\begin{enumerate}
 \item $D_1$ defined by $\boldq_1=\tfrac{2\pi}{L}(-10,19, 7)$,
 \item $D_2$ defined by $\boldq_2=\tfrac{2\pi}{L}(-11,13,15)$, 
 \item $D_3$ defined by $\boldq_3=\tfrac{2\pi}{L}( -4, 8,21)$, 
 \item $D_4$ defined by $\boldq_4=\tfrac{2\pi}{L}(-14,15,10)$, 
 \item $D_5$ defined by $\boldq_5=\tfrac{2\pi}{L}(-16, 8,14)$,
 \item $D_6$ defined by $\boldq_6=\tfrac{2\pi}{L}(-13,18,-5)$,
 \item $D_0$ defined by nucleons that are not in $D_i$, $i=1,\hdots,6$.
\end{enumerate}
Here $L=403\unit{fm}$ is the length of the box. 
We chose the domains ordered by the values of $S^e(\boldq)$ in the final 
configuration omitting angles within $15\degree$ of $\boldq_i$, $i=1,\hdots,6$.
We notice that domain $D_1$ does not coincide with the domain which occupies the 
largest volume by the end of the simulation, which is domain $D_2$. 
This may be due nucleons in domain $D_1$ having less deviation from their 
average position than nucleons in domain $D_2$.

\begin{figure}[!htb]
\centering
\includegraphics[trim=20 0 20 0, clip, width=0.45\textwidth] 
{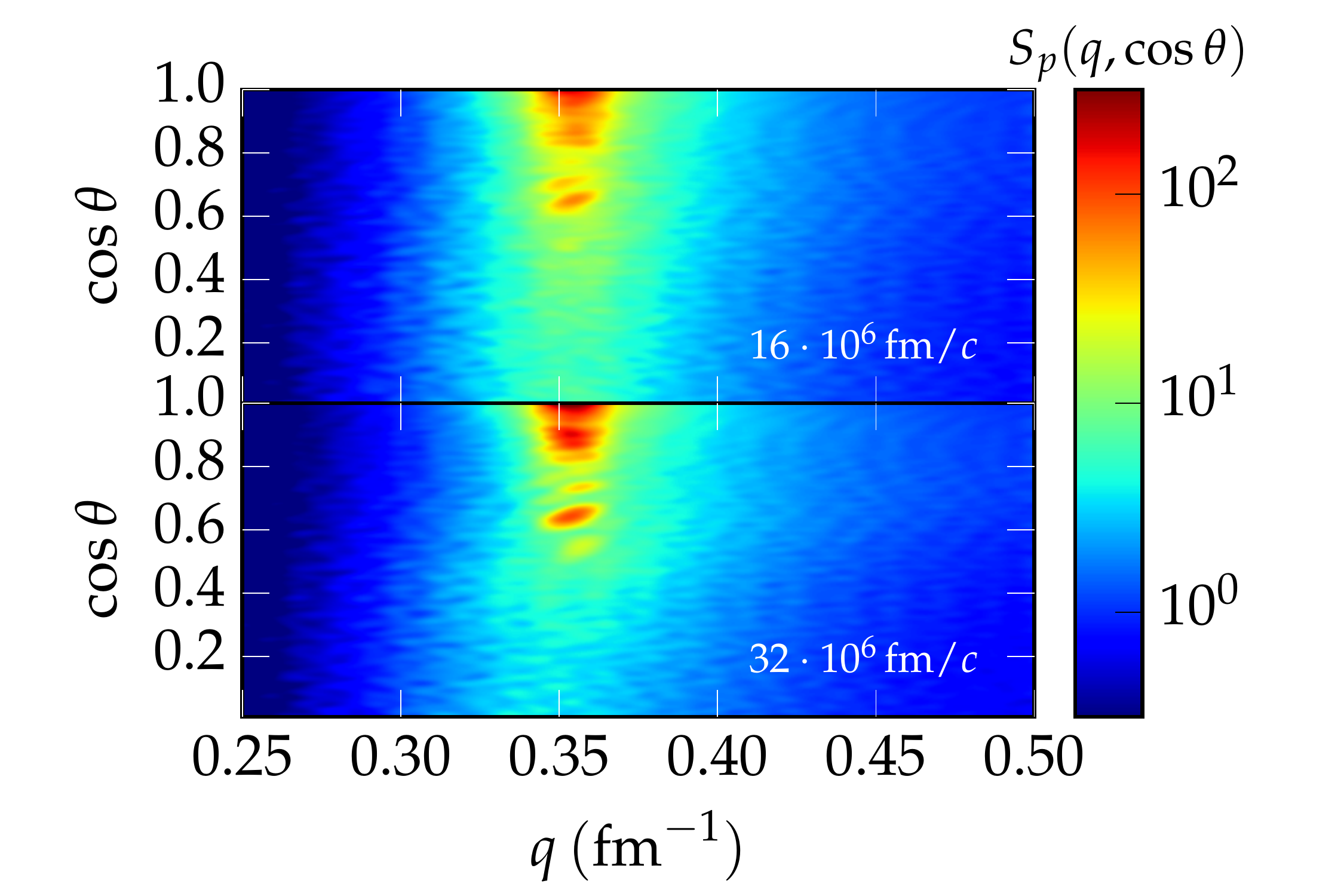}
\caption{\label{fig:sq_3276800} (Color online) Similar to Fig. 
\ref{fig:sq_0819200} but for the3\,276\,800  nucleon system at times $t=16$ 
and $32\cdot10^6\unit{fm}/c$.}
\end{figure}

From the structure factor plot, Fig. \ref{fig:sq_3276800}, we also see that 
this simulation has multiple large domains at the end of the run. 
This is clear from the existence of a large area with $S_p(q,\cos\theta) 
\gtrsim 10^2$ around $q\sim0.35\unit{fm}^{-1}$ instead of one or 
two localized peaks like in the smaller simulations. 
Over time the magnitude of $S_p(q,\cos\theta)$ increases for 
$q\sim0.35\unit{fm}^{-1}$ and $\cos\theta\gtrsim0.6$ while decreasing for 
$\cos\theta\lesssim0.6$. 
This follows from the defect domain $D_0$, which includes small domains that 
form an angle $\theta\gtrsim45\degree$ with respect to domain $D_1$, decreasing 
from 40\% in volume to 15\% from $t=16$ to $t=32\cdot10^6\unit{fm}/c$.

\begin{figure*}[!htb]
\centering
\includegraphics[width=0.8\textwidth]{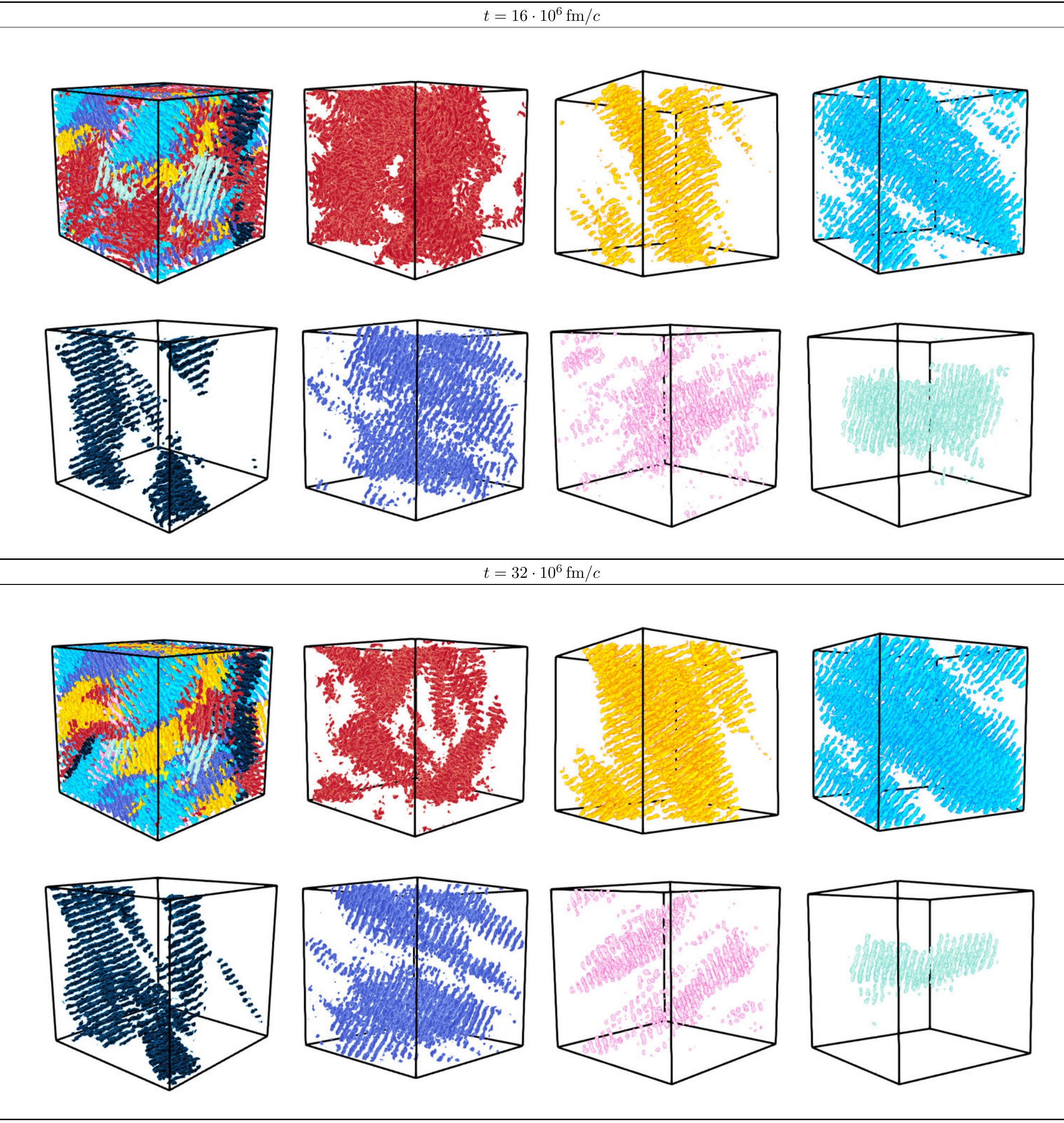}
\caption{\label{fig:3276800} (Color online) Configuration of our 3\,276\,800 
nucleon simulation at two different times, $t=16\cdot10^6\unit{fm}/c$ and 
$t=32\cdot10^6\unit{fm}/c$.  
We show six different domains in our run with nucleons: $D_0$ (red), $D_1$ 
(yellow), $D_2$ (light blue), $D_3$ (black), $D_4$ (dark blue), $D_5$ (pink), 
and $D_6$ (). 
In the top row of each time we show from left to right all domains in the system 
followed by domains $D_0$, $D_1$, and $D_2$.
In the bottom row from left to right we show domains $D_3$, $D_4$, $D_5$, and 
$D_6$.}
\end{figure*}

The matrix of the angles between the 6 largest domains (all domains chosen not 
including the ones that make up $D_0$) is given by 
\begin{equation}
\frac{\boldq_i\cdot\boldq_j}
{\vert\boldq_i\vert\vert\boldq_j\vert}
=\left(
\begin{array}{rrrrrr}
 0.0\degree& 25.6\degree& 48.9\degree& 16.2\degree& 36.9\degree& 31.8\degree\\
25.6\degree&  0.0\degree& 26.6\degree& 15.6\degree& 18.1\degree& 54.2\degree\\
48.9\degree& 26.6\degree&  0.0\degree& 42.2\degree& 35.5\degree& 79.9\degree\\
16.2\degree& 15.6\degree& 42.2\degree&  0.0\degree& 21.0\degree& 39.3\degree\\
36.9\degree& 18.1\degree& 35.5\degree& 21.0\degree&  0.0\degree& 56.9\degree\\
31.8\degree& 54.2\degree& 79.9\degree& 39.3\degree& 56.9\degree&  0.0\degree
\end{array}
\right).
\end{equation}
As observed for the two main domains in the smaller simulations, the system is 
dominated by domains that form angles $\theta\lesssim45\degree$ with each 
other. 
Only domain $D_6$ is consistently found at angles $\theta \gtrsim 45\degree$ 
with respect to other domains. 
As shown in Figs. \ref{fig:e_3276800} and \ref{fig:3276800}, it has a volume 
similar to domain $D_1$ halfway through the simulation but almost disappears by 
the end of the run. 
This seems to indicate that for the ``waffle'' phase domains nearly 
perpendicular to other ones disappear first, likely due to the large energy 
that need to be stored in its interface with other domains. 
The defects domain $D_0$ also decreases considerably in volume by 
the end of the run when compared to the halfway point. 
Most of its volume was absorbed by the domains $D_i$, $i=1,\hdots,4$.

Due to the high computational cost of this run we do not evolve it any further. 
Based on the results for the other simulations we speculate that if run for 
longer all domains in this simulation will eventually converge to a single one.
It is unclear, though, which one of the four larger domains at the end of the 
run would prevail over the others or if any other domains would appear.

\subsection{Simulations with $Y_p=0.40$}\label{ssec:yp40}

We examine simulations of seven different sizes for MD simulations with 
proton fraction $Y_p=0.40$. 
Five runs were already discussed in Ref. \cite{schneider:16}; the ones
containing 51\,200, 76\,800, 102\,400, 204\,800, and 409\,600 nucleons that 
were evolved at $n=0.05\unit{fm}^{-3}$ at $T=1\unit{MeV}$ for at least $10 
\cdot 10^6\unit{fm}/c$. 
The run with 409\,600 nucleons was evolved for a further $3\cdot10^6\unit{fm}/c$ 
for this work as its defects were not fully equilibrated. 
This has little effect on our estimate for the impurity parameter $Q_{\rm imp}$ 
of the pasta, our main result in Ref. \cite{schneider:16}. 
We include two additional runs: a small one with 61\,440 nucleons and a 
large one with 819\,200 nucleons. 
If let to evolve without the influence of any external potentials all of these 
systems form plates connected by Terasaki ramps \cite{terasaki:13, horowitz:15, 
guven:14, berry:16, schneider:16}. 
A summary of these runs is discussed in Table \ref{Tab:yp40}.

\begin{table}[htb!]
\caption{\label{Tab:yp40} Summary of our MD runs with proton fraction 
$Y_p=0.40$. We list the number of nucleons in the first column, the total 
evolution time in the second column and the side of the simulation box in the 
third column. In the fourth and fifth columns we enumerate, respectively, the 
number of left-handed and right-handed Terasaki ramps. In the sixth column we 
describe the ramps configuration, see text and Fig. \ref{fig:yp40}.}
\begin{ruledtabular}
\begin{tabular}{*{1}{r} c c c c c}
Nucleons & $t_{\mathrm total}$ & $L_{\mathrm box}$ 
& Left & Right & Configuration \\
& ($10^6$ fm/$c$) & (fm) & & & \\
\hline
      51\,200  &   10.0  & 100.8 & 4 & 4 & dipole    \\
      61\,440  &   13.5  & 107.1 & 4 & 4 & dipole    \\
      76\,800  &   14.5  & 115.4 & 2 & 0 & isolated  \\
     102\,400  &   12.0  & 127.0 & 4 & 4 & dipole    \\
     204\,800  &   18.0  & 160.0 & 1 & 1 & dipole    \\
     409\,600  &   17.0  & 201.6 & 1 & 1 & dipole    \\
     819\,200  &   18.0  & 254.0 & 1 & 1 & dipole    \\
\end{tabular}
\end{ruledtabular}
\end{table}

We also perform runs of the same seven sizes acted upon by an external 
sinusoidal potential following Ref. \cite{schneider:16}. 
The external potential is removed after a short time, $0.1 \cdot 
10^6\unit{fm}/c$, and the runs are left to equilibrate for another $2.9 \cdot 
10^6\unit{fm}/c$. 
Due to the initial influence of the external potential, parallel plates form. 
In all cases, the parallel plates are only stable for runs with the number of 
plates detailed in Tab. \ref{Tab:plates}. 
When trying to create a different number of parallel plates within the 
simulation volume the plates quickly became unstable after the removal of 
the external potential and merge to form defects. 
We did not study the topology evolution of runs where unstable parallel plates 
merged after a short simulation time, even though that may be an interesting 
problem on its own.

\begin{table}[htb!]
\caption{\label{Tab:plates} Number of plates $N_p$ and distance $d$ between 
the center of neighboring plates for simulations of different sizes. Runs 
marked with a $\dagger$ were performed for a previous work \cite{schneider:16} 
while $\ddagger$ denotes new runs. }
\begin{ruledtabular}
\begin{tabular}{*{3}{r}}
     Nucleons   &  $N_p$  & $d$ (fm) \\
\hline
  \,\,51\,200$\dagger$ &  \,\,6  &   16.8  \\
  \,\,61\,440$\ddagger$&  \,\,6  &   17.9  \\
  \,\,76\,800$\dagger$ &  \,\,7  &   16.5  \\
     102\,400$\dagger$ &  \,\,7  &   18.1  \\
     204\,800$\dagger$ &     10  &   16.0  \\
     409\,600$\dagger$ &     11  &   18.3  \\
     812\,900$\ddagger$&     14  &   18.1  \\
\end{tabular}
\end{ruledtabular}
\end{table}

\begin{figure}[!htb]
\centering
\includegraphics[trim=0 10 0 0, clip, width=0.45\textwidth] 
{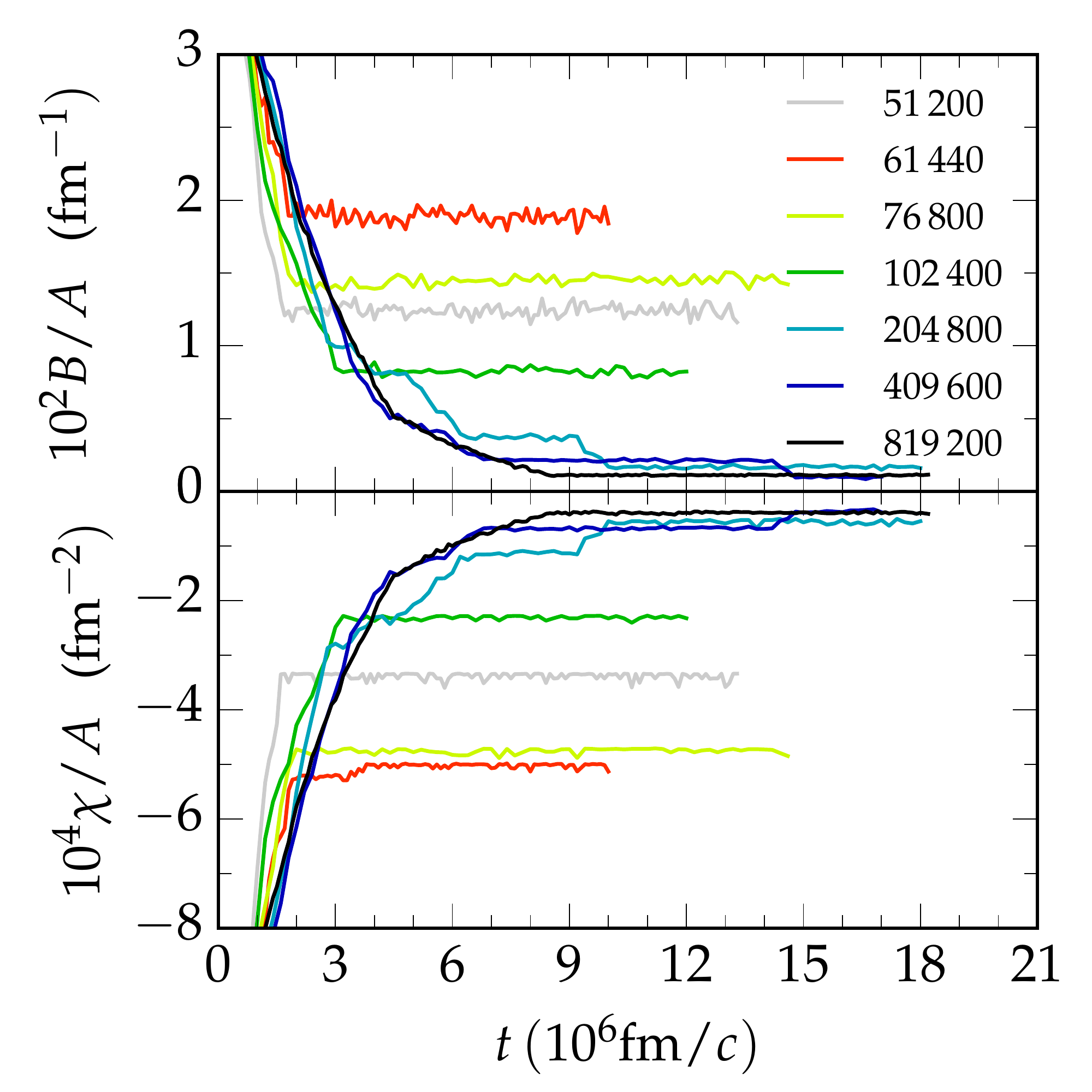}
\caption{\label{fig:M_40} (Color online) Plots of the normalized mean curvature 
$B/A$ (top) and normalized mean Gaussian curvature $\chi/A$ (bottom) as a 
function of simulation time $t$ for seven simulations with $Y_p=0.40$,  
$n=0.050\unit{fm}^{-3}$ and $T=1.0$ MeV.}
\end{figure}

In Fig. \ref{fig:M_40} we show the normalized mean curvature and normalized 
Gaussian curvatures for the $Y_p=0.40$ simulations \cite{schneider:13, 
schneider:14}. 
The four smaller runs seemingly converged to a stable configuration within 
$3\cdot10^6\unit{fm}/c$, while the larger ones took four to five times longer. 
Note that the 51\,200, 409\,600, and 819\,200 nucleon systems with $Y_p=0.40$ 
have equilibrated in, respectively, $2$, $15$, and $9\cdot10^6\unit{fm}/c$. 
These time scales are significantly faster than the convergence time for 
$Y_p=0.30$ runs of the same size. 
This is valuable as the computational cost of a run scales with 
$\mathcal{O}(N^2Y_p^2)$. 
Furthermore, the three larger simulations have very similar curvatures at the 
end of the runs, while the four smaller ones do not seem to obey any clear 
trend with respect to their size. 
As we will show below this is due to the types of defects formed in each of 
the runs.

Similarly to the $Y_p=0.30$ case we use our algorithm of Sec. \ref{ssec:domains} 
to separate the system in different domains. 
In the $Y_p=0.40$ cases, however, we only analyze two domains. 
Domain $D_1$ is defined by protons with structure factor $S_j^e(\boldq_{\rm 
max},t_f) > 0.40$, see Eq. \eqref{eq:seq}.
Here $\boldq_{\rm max}$ is the most common normal to the plates formed in each 
system and obtained from the highest peak in $S(\boldq)$, shown in Fig. 
\ref{fig:sq_40}. 
Protons which do not belong to domain $D_1$ are set as part of domain $D_0$.

\begin{figure}[!htb]
\centering
\includegraphics[trim=0 10 0 0, clip, width=0.45\textwidth] 
{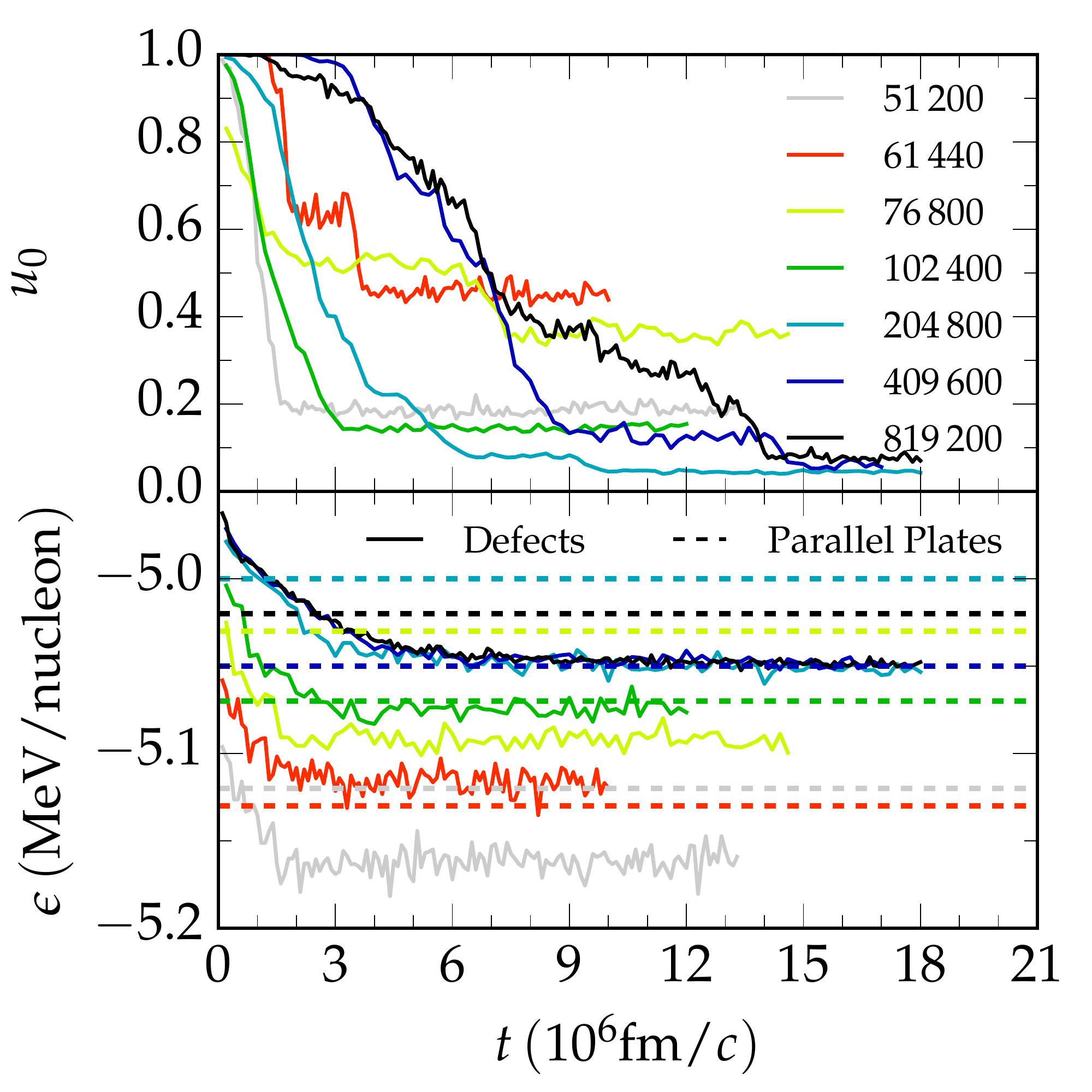}
\caption{\label{fig:V_40} (Color online) Volume fraction $u_0$ of nucleons in 
domain $D_0$ (top) and potential energy per nucleon $\epsilon$ (bottom). 
Domain $D_0$ is formed by defects while $D_1$ defined by parallel plates 
perpendicular to $\boldq_1=\boldq_{\rm max}$, see text. 
Except for the 76\,800 run domain $D_1$ is formed exclusively by parallel 
plates, and thus $u_0$ is the volume of defects.}
\end{figure}

\begin{figure}[!htb]
\centering
\includegraphics[trim=5 30 0 0, clip, width=0.45\textwidth]{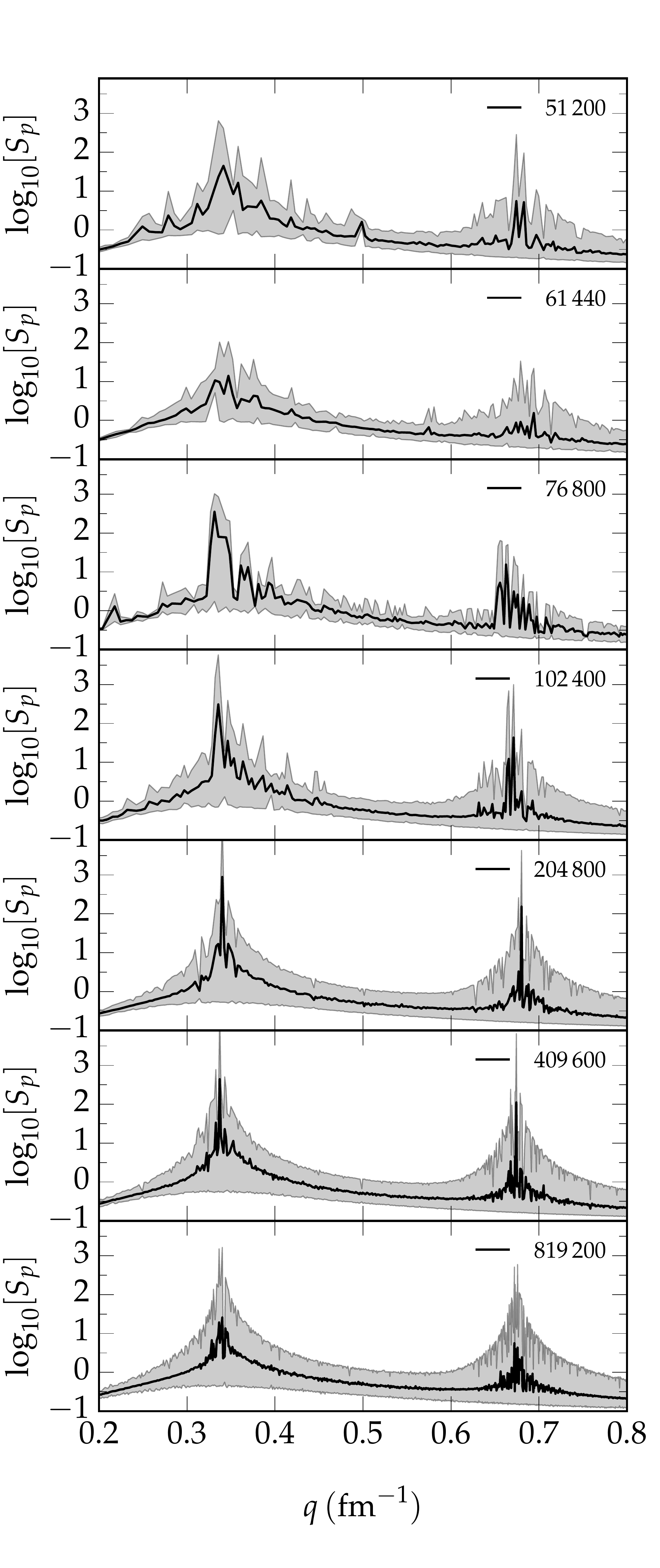}
\caption{\label{fig:sq_40} Plots of the angle averaged proton structure factor 
$S_p(q)=\langle{S_p(\boldq)}\rangle_{q}$ (thick black lines) for the 
last $1.0\cdot10^6\unit{fm}/c$ of each simulation run. The average is bounded 
by the maximum and minimum in $S_p(q)$ for each $q=\vert\boldq\vert$ 
(shaded grey area).}
\end{figure}

In the top panel of Fig. \ref{fig:V_40} we plot the volume fraction $u_0$ of 
nucleons in domain $D_0$ (top) for the runs with defects.  
The volume occupied by domain $D_1$ is $u_1=1-u_0$. 
The three larger simulations have, at the end of their run, a very similar 
volume fraction of defects, alluding that topology and defect density may have 
converged for the larger runs. 
As in the curvature case, the smaller runs do not show any clear trend with 
respect to their size. 
However, the absolute value of curvatures do seem correlated amongst themselves 
and with the volume fraction $u_0$ occupied by the defects domain $D_0$.

In the bottom panel of Fig. \ref{fig:V_40} we plot the energy per nucleon 
$\epsilon$ of the systems with defects and compare with the systems forced to 
form parallel plates perpendicular to one of the sides of the box by an 
external potential. 
For most simulation sizes the energy per nucleon is lower for systems that have 
defects instead of parallel plates. 
The exceptions are the runs with 61\,440 that has a larger energy per 
nucleon in the system with defects, and the runs with 102\,400, and 409\,600, 
where the energies are almost the same in both cases. 
Ideally, we expect a system with parallel plates to have smaller energy per 
particle than one with defects.  
Our results showing that often to be otherwise is a consequence of finite size 
effects of the systems. 
Slow expansion runs with up to 102\,400 nucleons similar to the ones of 
Schneider \etal \cite{schneider:13} show that parallel plates tilted with 
respect to the sides of the box can form at $n=0.05\unit{fm}^{-3}$. 
These tilted plates have lower energies per particle than the ones obtained for 
either the run with defects and the ones with parallel plates discussed here.

\begin{figure}[!htb]
\centering
\includegraphics[trim=5 30 0 0, clip, width = 0.45\textwidth] 
{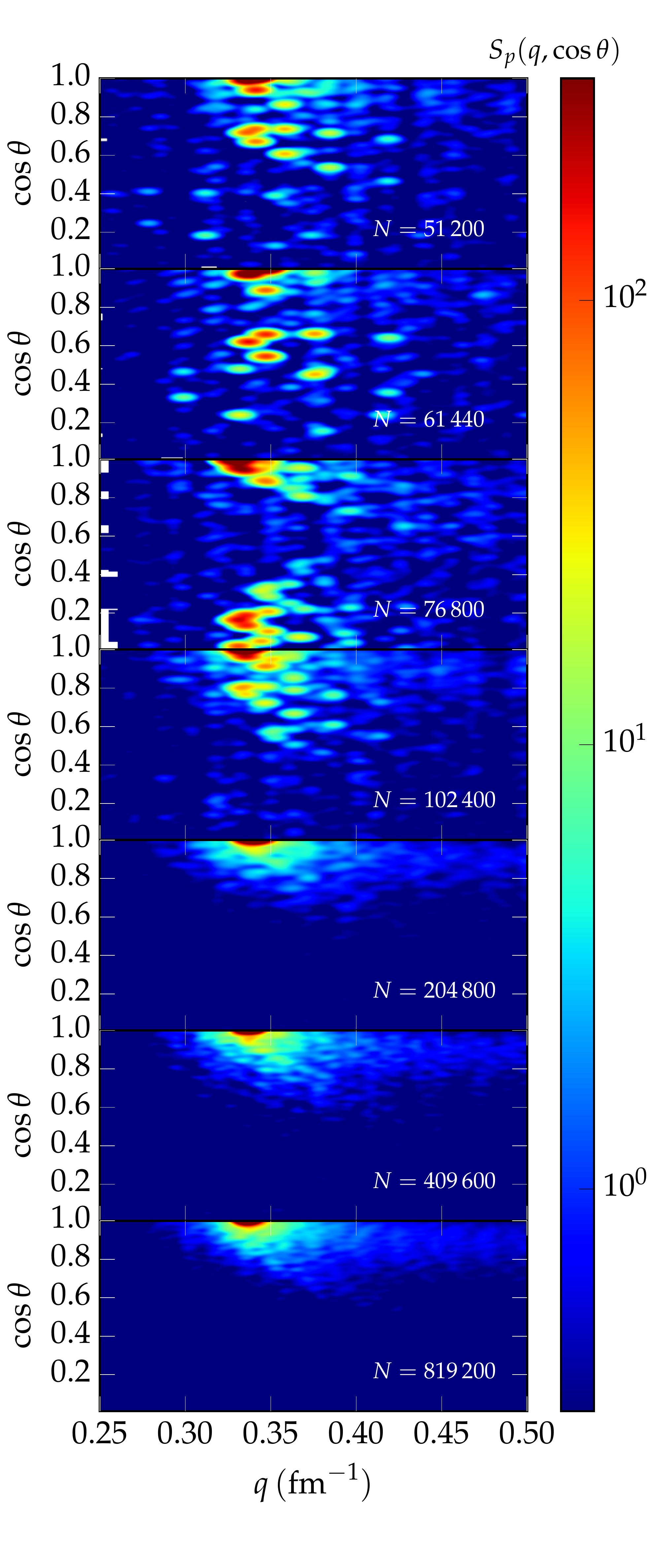}
\caption{\label{fig:sqx_40} Azimuthal average $S_p(q,\cos\theta)$ of the proton 
structure factor for the last $1.0\cdot10^6\unit{fm}/c$ of each run. 
The angle $\theta$ is defined in Eq. \eqref{eq:theta}.}
\end{figure}

In Fig. \ref{fig:sq_40} we show the angle average structure factor for 
protons $S_p(q)$ for our seven simulations as well as their upper and lower 
bounds, defined by the maxima and the minima in $S_p(\boldq)$ for a given 
$q=\vert\boldq\vert$. 
All structure factors have a similar qualitative behavior, with sharp 
peaks at $q'\sim0.34\unit{fm}^{-1}$ and $2q'$. 
The quantitative behavior, on the other hand, only seems to agree for the 
three larger simulations as the four smaller ones have a few other minor peaks 
between $q'$ and $2q'$ that don't appear in the larger ones. 
As we will show next, this is due to the different structures of the defects 
formed within the simulation volume.

The topology of the defects formed can be inferred from Fig. \ref{fig:sqx_40}, 
where we plot the structure factors $S_p(q,\cos\theta)$ with respect to the 
direction of $\boldq_{\rm max}$ where $S_p^e(\boldq)$ is a maximum. 
As in Sec. \ref{ssec:yp30} we histogram the values of $S_p(q,\cos\theta)$ and 
smooth it with a Gaussian filter. For better visualization we use standard 
deviations $\sigma_q=0.025\unit{fm}^{-1}$ and $\sigma_{\cos\theta}=0.05$ in 
the Gaussian filter for simulations with 204\,800 and larger and 
$\sigma_q=0.033\unit{fm}^{-1}$ and $\sigma_{\cos\theta}=0.067$ for simulations 
with 102\,400 nucleons or smaller. 
In the $S_p(q,\cos\theta)$ plots the main domain appears as a peak with 
$q\sim0.34\unit{fm}^{-1}$ and $\cos\theta\sim1$. 
Secondary domains appear as peaks with $q\sim0.34\unit{fm}^{-1}$ and 
$\cos\theta<0.9$. 
It is clear from these plots that the types of defects is different between the 
runs.

In Fig. \ref{fig:yp40} we show the final configurations for the $Y_p=0.40$ 
systems separated as two domains: $D_0$, defects, and $D_1$, defined by the 
maximum in $S(\boldq)$. 
With the exception of the 76\,800 nucleons simulation, domain $D_1$ is always
formed by parallel plates.

\begin{figure}[!htb]
\centering
\includegraphics[trim=0 -10 0 -10, clip, width = 0.4\textwidth] 
{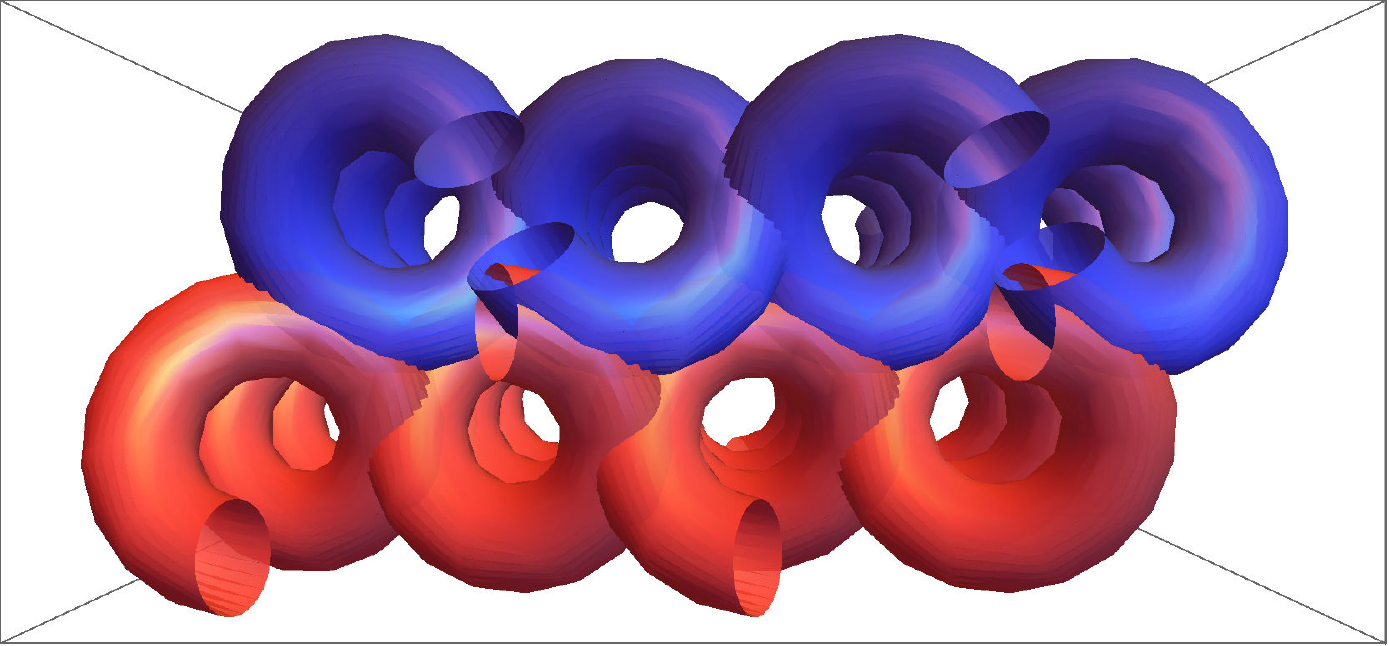}
\includegraphics[trim=0 -10 0 -10, clip, width = 0.4\textwidth] 
{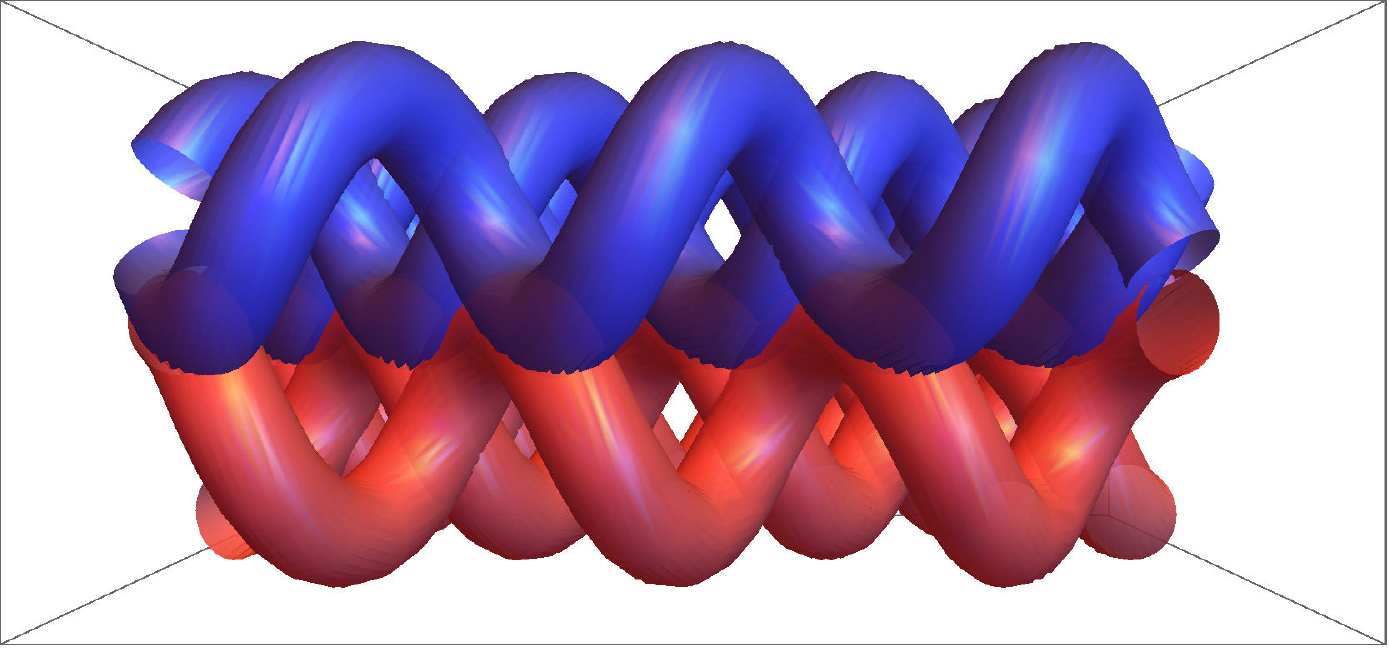}
\includegraphics[trim=0 -10 0 -10, clip, width = 0.4\textwidth, angle=90] 
{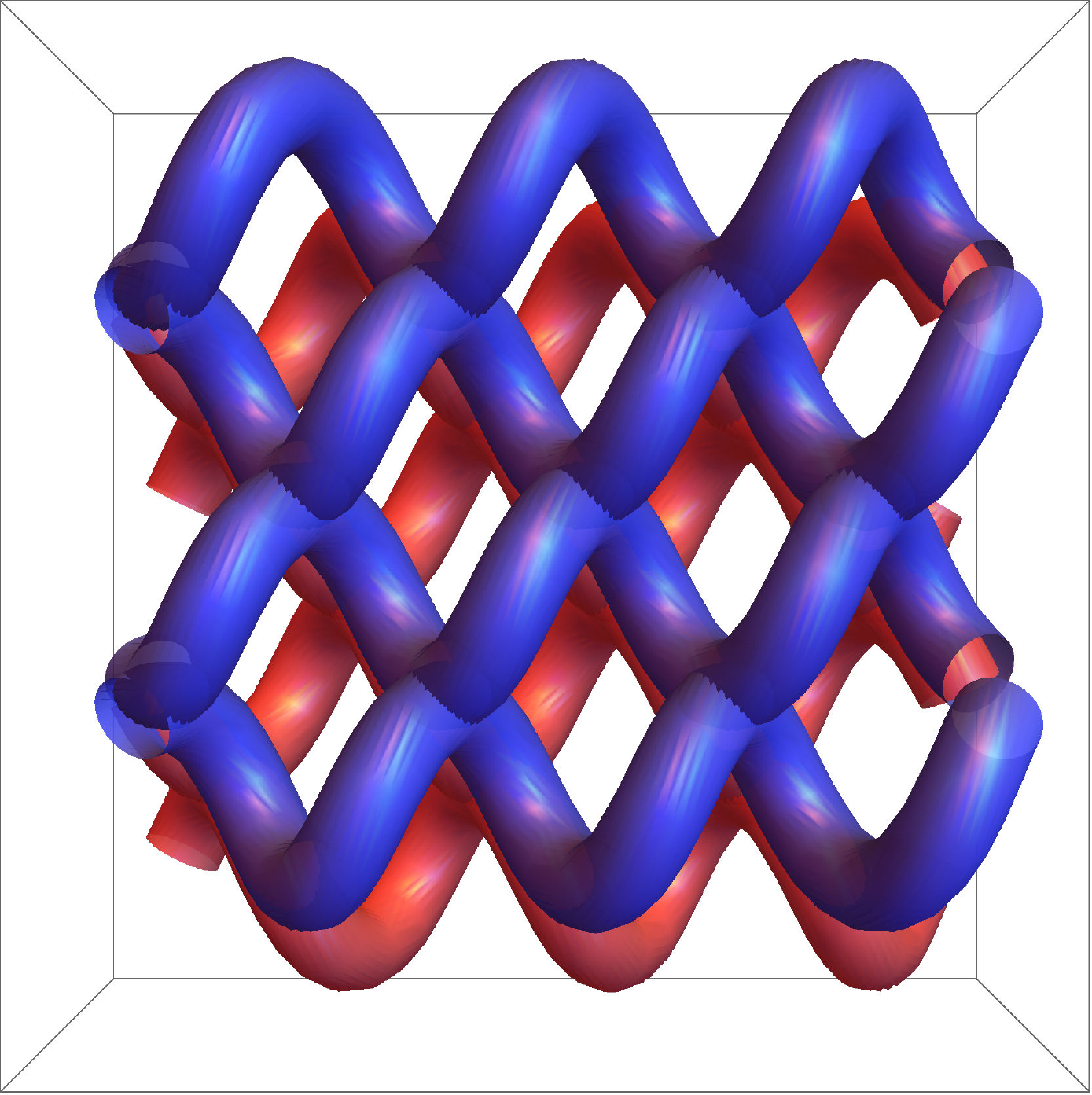}
\caption{\label{fig:8helices} Schematic top (top panel) and side views 
(center and bottom panels) of the dipole pattern formed by eight helical ramps. 
In red (blue) we show the right (left) -handed helices. In the bottom panel we 
can identify the position of the planes that would form a $45\degree$ angle 
with respect to the helices and connect the helices.}\end{figure}

In the 51\,200 nucleon system the normal to the plates and normal to the 
defects form an angle of about $45\degree$ with respect to each other. 
This is clear from the location of the second maxima in $S_p(q,\cos\theta)$ at
$q\sim 0.34\unit{fm}^{-1}$ and $\cos\theta \sim0.70$ ($\theta\sim 
40\degree$) seen in Fig. \ref{fig:sqx_40}. 
This is also clear from the configurations shown in the top row of Fig. 
\ref{fig:yp40}. 
The pattern of Terasaki ramps forms a dipole with eight helical ramps side by 
side, four left-handed and four right-handed helices, which connect the five
parallel plates within the simulation volume. 
This is the dipole pattern discussed in Refs. \cite{guven:14, berry:16}. 
In Fig. \ref{fig:8helices} we show a schematic picture of the 
defects since our domain detection algorithm does not clearly separate part of 
the helices from the planes in this case.

Similarly to the 51\,200 simulation, the 61\,440 nucleon run also forms a set 
of eight helices with the pattern scheme shown in Fig. \ref{fig:8helices}, see 
second row of Fig. \ref{fig:yp40}. 
The main difference here is that in this simulation the helices form in a 
different angle with the sides of our simulation box. 
In the 61\,440 nucleon case our domain algorithm performs better than in the 
51\,200 nucleon case and, thus, the volume fraction $u_0$ of domain $D_0$ 
appears to be twice the size in the slightly larger run as more protons are 
identified as belonging to domain $D_0$, see top panel of Fig. \ref{fig:V_40}.

The topology formed by the 76\,800 nucleon system is somewhat different than 
what we see in all other simulations. 
Here two sets of plates that are almost perpendicular to each other compete, 
with neither occupying significantly more than half of the simulation volume by 
the end of the run. 
This is seen by the location and magnitude of the second largest peak in 
$S_p(q,\cos\theta)$ which occurs at $q\sim 0.33\unit{fm}^{-1}$ and 
$\cos\theta\lesssim 0.2$ ($\theta\gtrsim78\degree$), Fig. \ref{fig:sqx_40}. 
This system is even more peculiar in that it formed two helical ramps 
perpendicular to each other, both of which are left-handed, see third row of 
Fig. \ref{fig:yp40}. 
This is unlike any of the other systems we have simulated where right- and 
left-handed ramps appear in equal numbers.

The 102\,400 nucleon system is very similar to the 51\,200 and 61\,440 systems: 
the helical ramps and plates are at an angle of approximately $45\degree$ 
with each other, Fig. \ref{fig:sqx_40}. 
However, the magnitude of the second peak in the 102\,400 nucleon system is 
smaller than in the 51\,200 nucleon system, fourth row in Fig. \ref{fig:yp40}, 
because in the larger system the defects occupy, proportionally, a smaller 
volume, see plot of $u_0$ in Fig. \ref{fig:V_40}.

\begin{figure}[!htb]
\centering
\includegraphics[trim=0 -10 0 -10, clip, width = 0.4\textwidth] 
{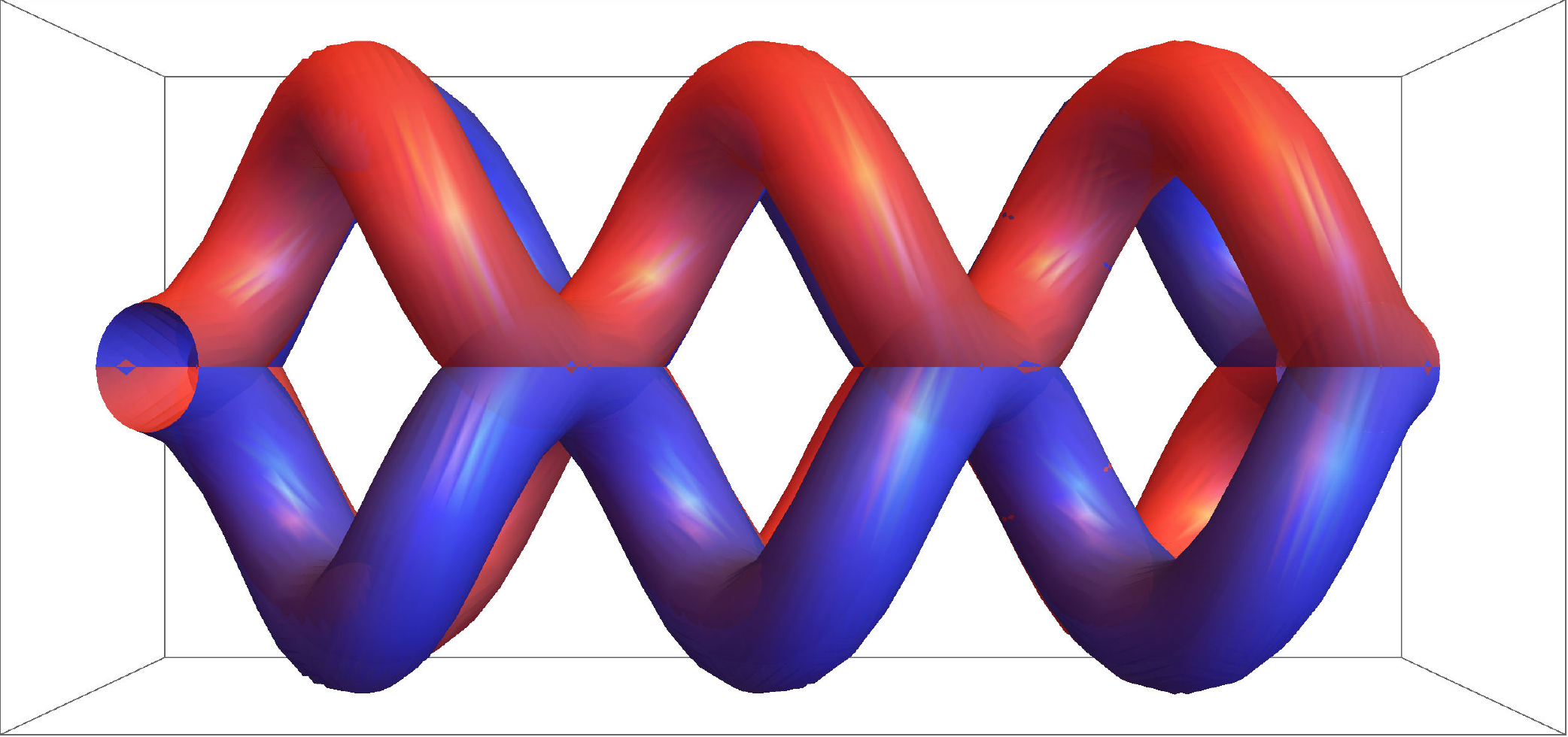}
\includegraphics[trim=0 -10 0 -10, clip, width = 0.4\textwidth] 
{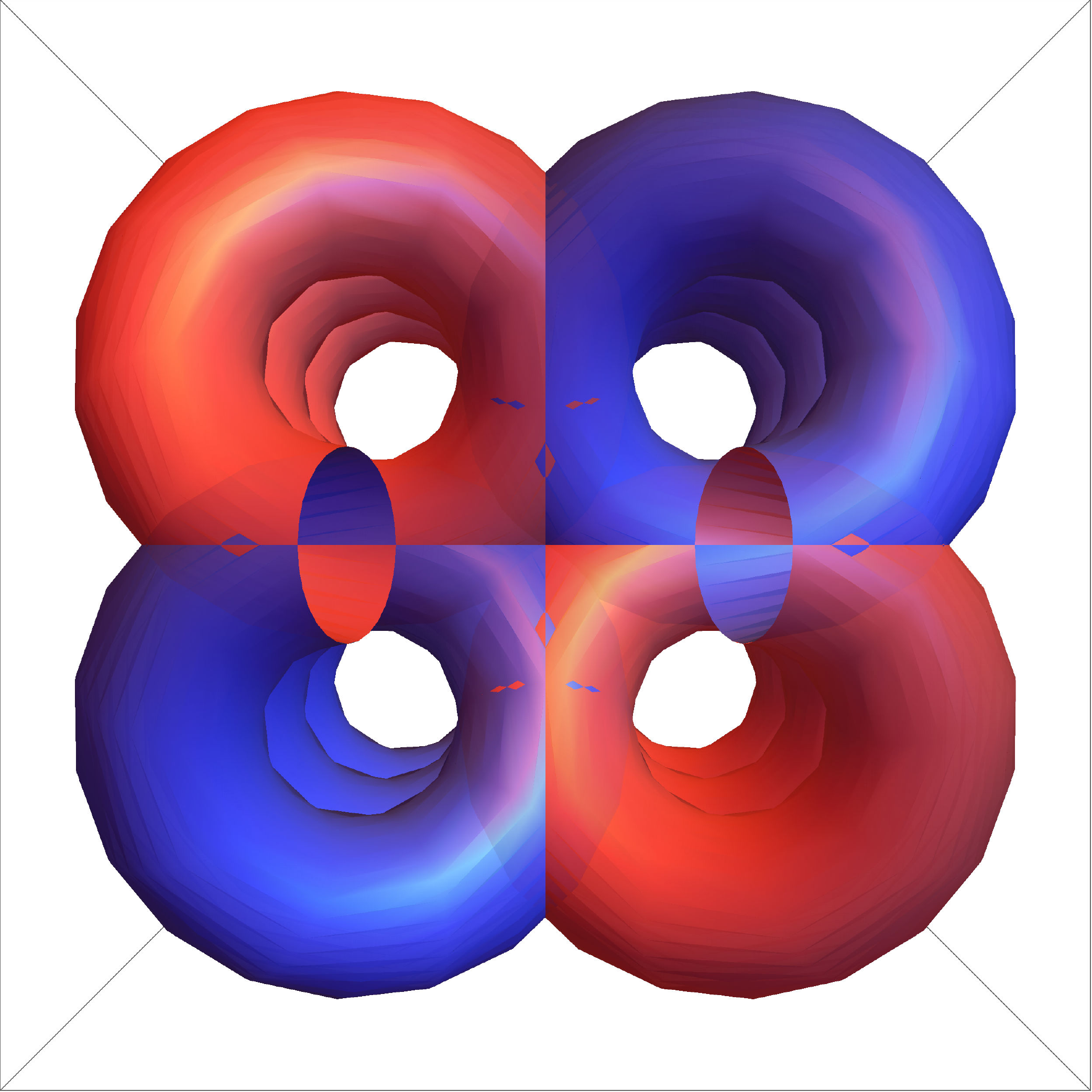}
\caption{\label{fig:4helices} Schematic side (top panel) and top views (bottom 
panel) of one of the quadrupole possibilities formed by four helical ramps. In 
red (blue) we show the right (left) -handed helices.}
\end{figure}

\begin{figure}[!htb]
\centering
\includegraphics[trim=0 -10 0 -10, clip, width = 0.4\textwidth] 
{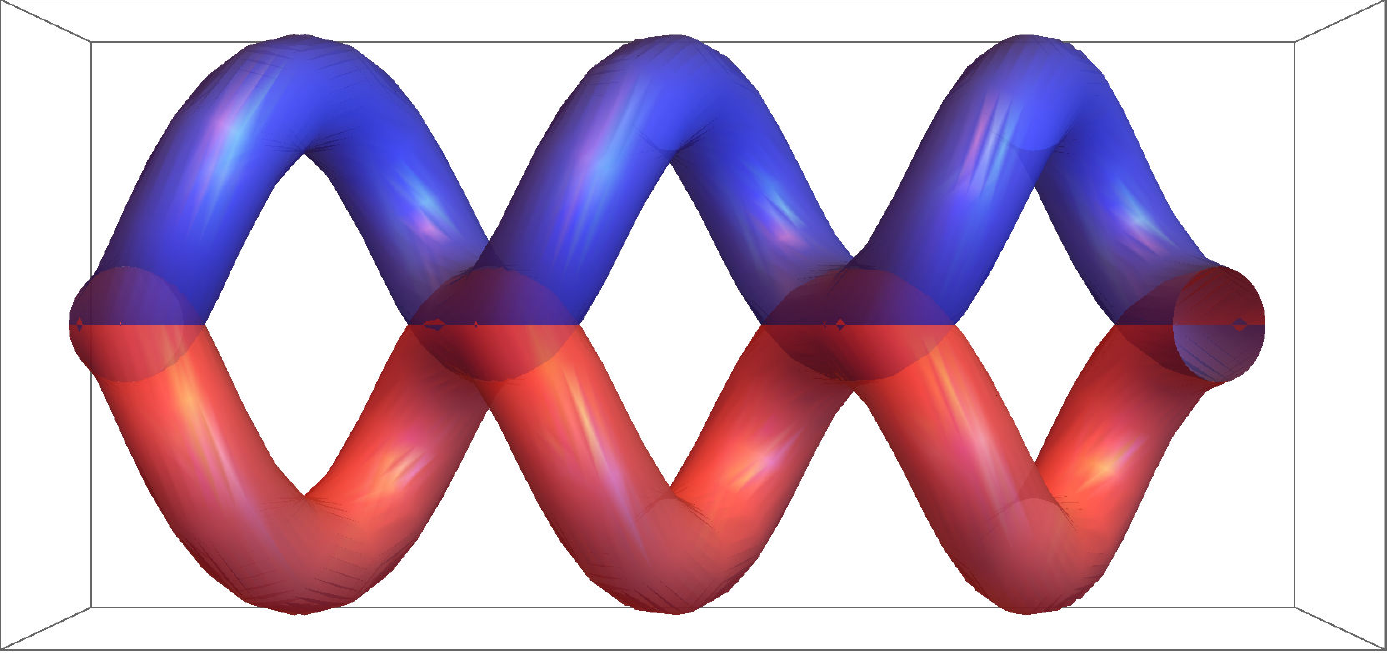}
\includegraphics[trim=0 -10 0 -10, clip, angle=90, width = 0.4\textwidth] 
{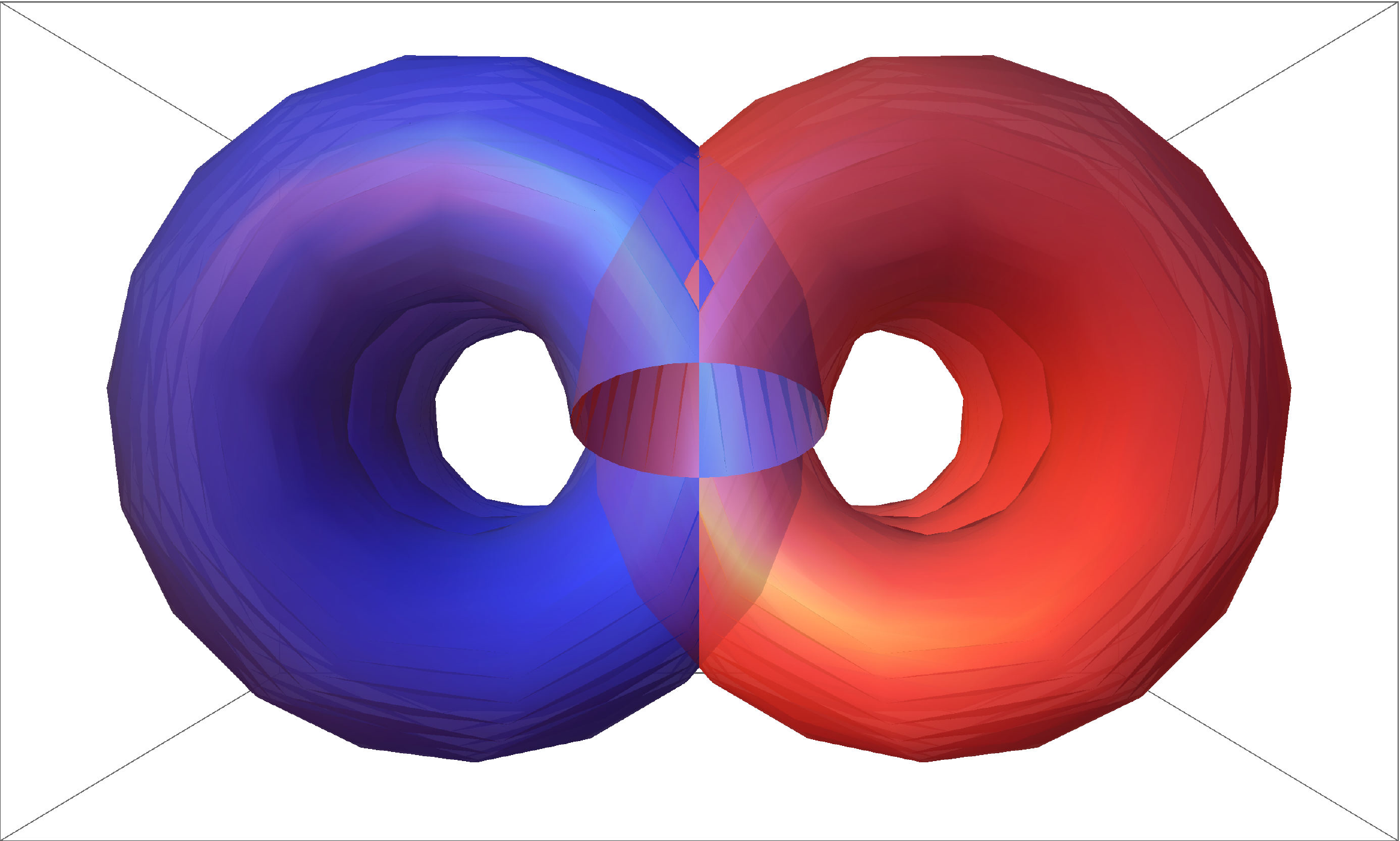}
\caption{\label{fig:2helices} Schematic side (top panel) and top views (bottom 
panel) of one of the dipole possibilities formed by two helical ramps. In red 
(blue) we show the right (left) -handed helices.}
\end{figure}

The three larger systems, with 204\,800, 409\,600, and 819\,200 nucleons, have 
a similar evolution history. 
Before achieving their final configuration, the three systems go through 
similar stages to the ones described by Berry \etal \cite{berry:16} and shown 
in their Fig. 1. 
However, due to the larger size of the simulations presented here, the system 
forms several ``ramps'' connecting its planes. 
Over time, ramps move towards each other and the ones with same helicity 
merge while pairs with opposite helicities persist. 
Pairs of ramps also attract each other as the system evolves. 
At this point, we speculate that two events can take place. 
The angle of approach of the pairs of helices can be such that it forms a 
quadrupole as the one schematically shown in Fig. \ref{fig:4helices}. 
This configure is stable and the system, likely, does not evolve further. 
This is what is observed by Berry \etal in their 75\,000 nucleon simulation 
\cite{berry:16}. 
In the large runs discussed here, however, the pairs of ramps approach each 
other in such a way that ramps with the same helicity face each other. 
As this happens, thermal fluctuations in the system cause ramps with same 
helicity to merge and a dipole as the one shown in Fig. \ref{fig:2helices} is 
created. 
Unlike the dipole configurations observed in the smaller runs, where helices 
and plates are at a $45\degree$ with each other, in the large runs the 
helices are at a $90\degree$ angle with the plates. 
This is also noticed by a lack of a second significant peak in 
$S_p(q,\cos\theta)$ in Fig. \ref{fig:sqx_40}.

\begin{figure}[!htb]
\centering
\includegraphics[width = 0.45\textwidth]{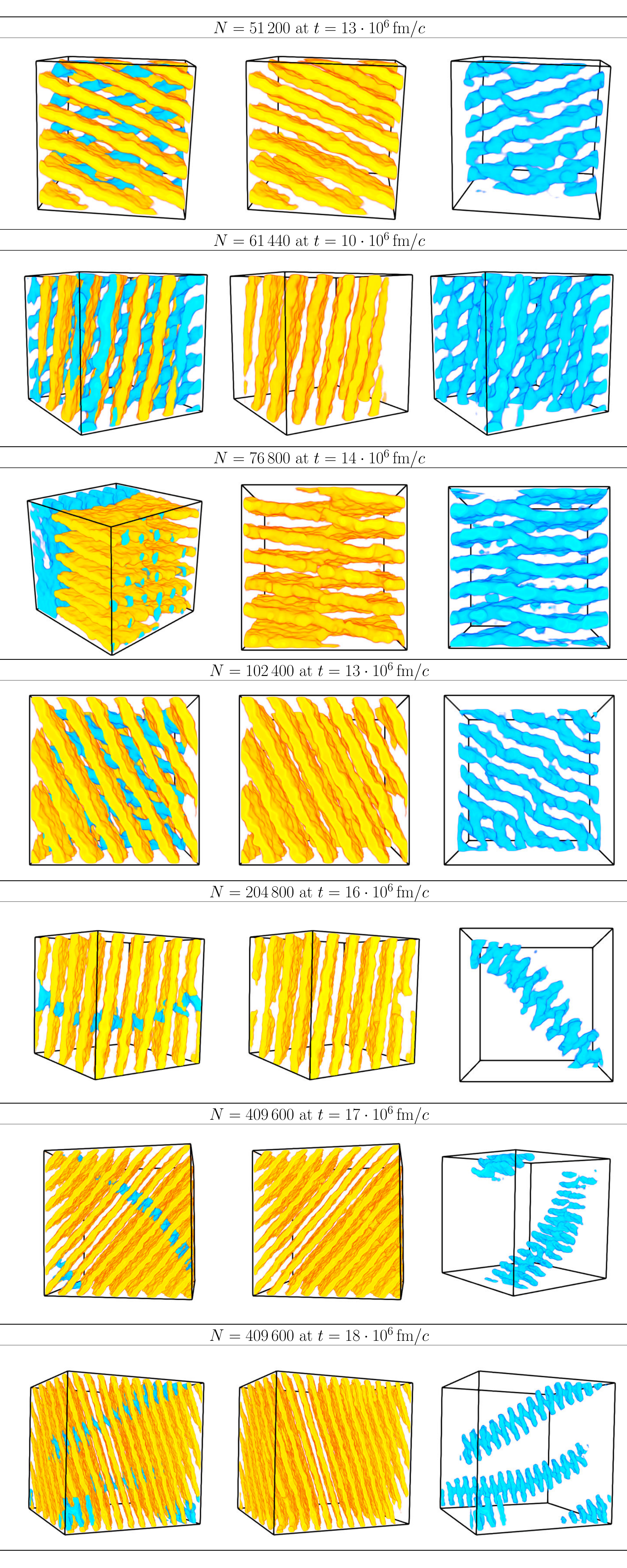}
\caption{\label{fig:yp40} (Color online) Last configuration of each of out 
$Y_p=0.40$ runs. For each system we show two different domains, $D_0$ (light 
blue) and $D_1$ (yellow).}
\end{figure}

Although both the curvatures and the structure factors in the $Y_p=0.40$ runs 
seem to have converged as the simulation size was increased it is unclear 
whether this convergence would remain true if larger systems were simulated. 
Furthermore, we observed three different types of defects in these simulations 
in the six runs performed. 
And yet another type was observed by Berry \etal \cite{berry:16} in similar MD 
simulations.
Presently, there is no clear way of knowing whether this would remain true if we 
simply repeated simulations for systems of the same size or if we performed 
even larger simulations. 
Thus, further studies of these phases are warranted.

\section{Conclusions}\label{sec:Conclusions}

Numerical simulations of nuclear pasta have attracted attention lately as 
we have finally reached a stage where efforts to find indirect evidence of its 
existence are underway. 
Nuclear matter properties at sub-saturation densities, where the pasta phases 
are likely to form, can be constrained from the cooling curves of accreting 
neutron stars in quiescence \cite{newton:13, merritt:16, brown:17, cumming:17, 
deibel:17} and from LIGO-Virgo combined searches for $r$-mode gravitational 
waves signals from spinning down neutron stars \cite{horowitz:12, pons:13, 
horowitz:15, passamonti:16}.  
It may also be the case that the neutrino signal from a galactic supernovae 
or a neutron star merger will shed light on the formation and properties of 
nuclear pasta \cite{horowitz:16, roggero:18}. 
Some of the pasta properties and its effects on physical observables are a 
function of the nucleon structure factors \cite{horowitz:04a, horowitz:04b}, 
which can be computed from numerical simulations. 
However, finite-size effects and computational limitations are a substantial 
problem that should be overcome in order to accurately determine nuclear pasta 
observables \cite{schneider:16}.

In this work we studied, to our knowledge, the largest nuclear pasta systems 
to date where nucleonic degrees of freedom are taken into account. 
Using the IUMD code and Big Red 2 and Titan computer resources we simulated 
nuclear pasta systems with up to 3\,276\,800 nucleons for proton fractions 
$Y_p=0.30$ and with up to 819\,200 nucleons for $Y_p=0.40$.

All $Y_p=0.30$ runs formed the expected ``waffle phase'' \cite{schneider:13, 
sagert:16}. 
We analyzed the structure factor dependence on simulation size and showed that 
there is qualitative agreement between the results obtained for simulations with 
51\,200 up to 3\,276\,800 nucleons. 
However, there are some quantitative differences in the results for 
simulations of such different sizes which are an artifact of the finite-size 
of the systems studied. 
Our results show that simulations with less than a hundred thousand nucleons 
still suffer from significant finite-size effects that need to be accurately 
addressed when predicting the transport properties of nuclear pasta, at least 
for the topologies studied in our work. 
Nonetheless, it is encouraging that there is a good agreement for the structure 
factor main peak location and its magnitude from much smaller simulations using 
a different method \cite{nandi:18}. 
Besides quantification of finite size effects, we introduced an algorithm that 
analyzes the evolution of domains within the simulations to test their formation 
and equilibration time-scales beyond what is possible by computing Minkowski 
functionals alone. 
We noticed that most of our $Y_p=0.30$ runs, if left to equilibrate for enough 
time, formed a single domain within the simulation volume. 
The exception being the largest of our runs, with 3\,276\,800 nucleons, which 
still had six large domains and many defects by the time we stopped evolving it 
due to its very high computational cost. 
The high cost of MD computations stemming from long range Coulomb 
repulsion between protons can be decreased with the implementation of robust 
fast multipole method algorithm for Yukawa-type potentials \cite{huang:09, 
zhang:10, baczewski:12}. 
Excluding significant advances in computer performance, this may be the 
only way to simulate nuclear pasta systems beyond a few million nucleons that 
need to be evolved for tens of millions of times steps in order to reach 
equilibrium.

We also performed a few MD simulations with proton fraction $Y_p=0.40$.
These runs, unless acted upon an external potential, formed parallel plates 
connected by ``Terasaki'' ramps  \cite{terasaki:13, horowitz:15, guven:14, 
berry:16, schneider:16}. 
For same size simulations, the $Y_p=0.40$ runs equilibrated significantly 
faster than their $Y_p=0.30$ counterparts. 
We found that the set of planes and Terasaki ramps formed different topologies 
that depended on simulation size. 
Amongst the topologies formed we observed dipoles composed of groups of eight 
parallel helical ramps, four left-handed and four right-handed, at an angle of 
$45\degree$ with the planes in three of our small simulations, the ones with 
51\,200, 61\,440, and 102\,400 nucleon runs. 
The three largest runs, in their final configuration, formed only one pair of 
parallel helical ramps, one left-handed and one right-handed. 
These ramps had a propensity to attract each other and form a dipole 
configuration at an angle of $90\degree$ with respect to the parallel planes. 
Finally, the simulation with 76\,800 nucleons was unique in that it formed two 
left-handed helices, and no right-handed ones. 
These helices were at $90\degree$ with respect to each other and with 
respect to the the planes they connected. 
We did not observe any quadrupole setup of helical ramps as seen by Berry \etal 
in Ref. \cite{berry:16} for a system with 75\,000 nucleons.
This may indicate that simulations with $\lesssim102\,400$ nucleons may be 
considerably sensitive to their size and the initial conditions of the 
simulation.

From the self-assembled patterns seen in our simulations and the time to 
establish and equilibrate them we estimate that, in order to minimize finite 
size-effects in computations of transport properties of nuclear pasta, it may be 
necessary to perform simulations with hundreds of thousands of nucleons. 
This is discouraging from the point of view of computational costs as 
simulations this large are unlikely to be possible anytime soon for full 
quantum-mechanical calculations \cite{sagert:16, fattoyev:17}. 
However, by understanding how finite-size effects affect the results 
for structure factor of nucleons and the transport properties of nuclear 
pasta we can make informed guesses about the direction which results should be 
corrected for smaller simulations, such as the ones shown here and by 
compilations of the results of Nandi and Schramm \cite{nandi:18}

\begin{acknowledgments}

We thank Greg Huber (KITP) and Kris Delaney (UCSB) for interesting and useful 
discussions regarding polymer topology and their similarities to nuclear pasta. 
We also thank William Newton (TAMU-Commerce) for sharing his insights on 
nuclear pasta domains, and Gerardo Ortiz (IU Bloomington) and Andrey Chugunov 
(IOFFE Institute St. Petersburg) for suggestions that helped improve this 
manuscript considerably.

A.\,S.\,S. was supported in part by the Conselho Nacional de Desenvolvimento 
Cient\'ifico e Tecnol\'ogico (201432/2014-5) and in part by the National Science 
Foundation under award No. AST-1333520 and CAREER PHY-1151197. 
M.\,E.\,C. is a Canadian Institute for Theoretical Astrophysics National Fellow.

This research was supported in part by DOE Grants No. DE-FG02-87ER40365 
(Indiana University) and No.  DE-SC0018083 (NUCLEI SciDAC-4 Collaboration), and 
in part by Lilly Endowment, Inc., through its support for the Indiana 
University Pervasive Technology Institute, and in part by the Indiana METACyt 
Initiative. The Indiana METACyt Initiative at IU is also supported in part by 
Lilly Endowment, Inc.

This research used resources of the Oak Ridge Leadership Computing Facility, 
which is a DOE Office of Science User Facility supported under Contract 
DE-AC05-00OR22725.

\end{acknowledgments}


\bibliography{large}

\end{document}